\documentclass[useAMS,usenatbib]{mn2e}
\usepackage{graphicx}

\title[Dynamical evolution of the Cybele asteroids]
{Dynamical evolution of the Cybele asteroids}
\author[V. Carruba, D. Nesvorn\'{y}, S. Aljbaae, and M. E. Huaman]{V. Carruba$^{1}$\thanks{E-mail: vcarruba@feg.unesp.br}, D. Nesvorn\'{y}$^{2}$, S. Aljbaae$^{1}$, and M. E. Huaman$^{1}$\\
$^{1}$UNESP, Univ. Estadual Paulista, Grupo de din\^{a}mica Orbital e
  Planetologia, Guaratinguet\'{a}, SP, 12516-410, Brazil \\
$^{2}$Department of Space Studies, Southwest Research Institute, Boulder, 
CO, 80302, USA\\
}

\begin{document}

\date{Accepted ... .  Received 2015 ...; in original form 2015 April 1}

\pagerange{\pageref{firstpage}--\pageref{lastpage}} \pubyear{2015}

\maketitle

\label{firstpage}

\begin{abstract}
The Cybele region, located between the 2J:-1A and 5J:-3A mean-motion 
resonances, is adjacent and exterior to the asteroid main belt.   An 
increasing density of three-body resonances makes the region between the 
Cybele and Hilda populations dynamically unstable, so that the Cybele zone 
could be considered the last outpost of an extended main belt.  The presence 
of binary asteroids with large primaries and small secondaries suggested that 
asteroid families should be found in this region, but only relatively 
recently the first dynamical groups were identified in this area.  Among 
these, the Sylvia group has been proposed to be one of the oldest families 
in the extended main belt. 

In this work we identify families in the Cybele region in the context of 
the local dynamics and non-gravitational forces such as the Yarkovsky and 
stochastic YORP effects.  We confirm the detection of the new Helga group 
at $\simeq$3.65~AU, that could extend the outer boundary of the Cybele 
region up to the 5J:-3A mean-motion resonance.  We obtain age estimates for 
the four families, Sylvia, Huberta, Ulla and Helga, currently detectable in 
the Cybele region, using Monte Carlo methods that include the effects of 
stochastic YORP and variability of the Solar luminosity. The Sylvia family 
should be $T = 1220 \pm 40$ Myr old, with a possible older secondary solution.
Any collisional Cybele group formed prior to the late heavy bombardment would 
have been most likely completely dispersed in the jumping Jupiter scenario of 
planetary migration.
\end{abstract}

\begin{keywords}
Minor planets, asteroids: general  Minor planets, asteroids: individual:
Cybele -- celestial mechanics.  
\end{keywords}
%

\section{Introduction}
\label{sec: intro}

The Cybele region, adjacent and exterior to the asteroid belt, is 
usually defined as the region in semi-major axis between the 2J:-1A and 5J:-3A 
mean-motion resonances with Jupiter.  An increase in the number density
of three-body resonances at semi-major axis larger than 3.7 (Gallardo 2014)
makes the region between the Cybele area and the Hilda asteroids dynamically 
unstable, making the Cybele region the last outpost of an ``extended'' main 
belt, according to some authors (Carruba et al. 2013).  Binary asteroids 
in the region suggested the presence of asteroid families, but only 
relatively recently a new dynamical family was found near 87 Sylvia 
(Vokrouhlick\'{y} et al. 2010).   The same authors
suggested that 107 Camilla and 121 Hermione could have had families in the past
that were dispersed by the local dynamics.   The possibility that the
Sylvia group might have been created just after the last phases of planetary
migration, 3.8 Gyr ago, suggested by Vokrouhlick\'{y} et al. 
(2010), may have interesting repercussions with regard to 
our understanding of the early Solar System history, and was one 
of the reasons why we started this research.

In this work we obtained dynamical groups in a newly identified domain 
of 1500 numbered and multi-opposition asteroids with high-quality 
synthetic proper elements, and revised the dynamical and physical properties.
We then obtain age estimates of the four families identified in this work,
Sylvia, Huberta, Ulla and Helga, with the method of Yarkovsky isolines 
and the Monte Carlo Yarko-Yorp approach of Vokrouhlick\'{y} et al. 
(2006a, b, c).  This method was modified to account for the stochastic 
YORP effect of Bottke et al. (2015), and historical changes 
of the Solar luminosity.  Past and future dynamically evolution of 
present and past families is investigated using new symplectic integrators 
developed for this work.  Finally, we analyzed the dynamical evolution of a 
fictitious Sylvia family formed before the late heavy bombardment in the 
jumping Jupiter scenario (Case I) of Nesvorn\'{y} et al. (2013).  We confirm 
the presence of the new Helga group (Vinogradova and Shor 2014) that could 
extend the boundary of the Cybele region up to the 5J:-3A mean-motion 
resonance, and the fact that the estimated age of the Sylvia could be 
compatible with an origin just prior to the late heavy bombardment.  
Families around 107 Camilla and 121 Hermione should disperse in timescales 
of the order of 1~Gyr, and could not be currently detectable.

This paper is so divided: in Sect.~\ref{sec: prop_el} we obtain synthetic
proper elements for numbered and multi-opposition asteroids in the Cybele
region, and review properties of mean-motion resonances in the area.
In Sect.~\ref{sec: sec_dyn} we revise the role that secular dynamics
play locally, and identify the population of asteroids currently in 
librating states of secular resonances. Sect.~\ref{sec: taxonomy}
deals with revising physical properties, when available, for local
objects, while Sect.~\ref{sec: fam_ide} discusses the problem of
family identification, dynamical interlopers, and preliminary chronology 
using the method of Yarkovsky isolines.  In Sect.~\ref{sec: chron} we 
obtain more refined estimates of the family ages and ejection velocities
parameters using the Yarko-Yorp approach of Vokrouhlick\'{y} et al.
(2006a, b, c) modified to account for the stochastic YORP effect of Bottke
et al. (2015), and variability of the Solar luminosity.  
Sect.~\ref{sec: dyn_cybele} deals with investigating past and future dynamical 
evolution of present and (possibly) past families in the area using newly 
developed symplectic integrators that simulates both the stochastic YORP 
effect and variances in the past values of the Solar luminosity.  
Sect.~\ref{sec: syl_LHB} deals with the dynamical evolution of a fictitious 
Sylvia family formed before Jupiter jumped in the case I scenario of the
jumping Jupiter model of Nesvorn\'{y} et al. (2013).  Finally, in 
Sect.~\ref{sec: conc}, we present our conclusions.

\section{Proper elements}
\label{sec: prop_el}

We start our analysis by obtaining proper elements for asteroids
in the Cybele region (defined as the region between the 2J:-1A and
5J:-3A mean-motion resonances with Jupiter, i.e., with an osculating 
semi-major axis between 3.28 and 3.70~AU).   We computed
synthetic proper elements for the 1507 numbered
and 433 multi-opposition asteroids in the region, whose osculating
elements were downloaded from the AstDyS site 
(http://hamilton.dm.unipi.it/astdys,
Kne\v{z}evi\'{c} and Milani 2003) on October 29th 2014, with the procedure
discussed in Carruba (2010b). Of the 1940 particles, 1327 numbered
asteroids and 317 multi-opposition survived for the length of the 
integration (10 Myr). We also eliminated all objects for which one of 
the proper elements $a,e, \sin{(i)}$ or proper frequencies $g$ and $s$ had 
errors larger than those classified as 
``pathological'' by Kne\v{z}evi\'c and Milani (2003); i.e., 
$\Delta a = 0.01$~AU, $\Delta e = 0.1$, $\Delta \sin{(i)} = 0.03$, 
and $\Delta g = \Delta s = 10~arc-sec \cdot yr^{-1}$.   This left us
with a data-set of 1225 numbered and 275 multi-opposition 
asteroids with proper elements in the orbital region, for a total
of 1500 asteroids with good-quality proper elements.

\begin{figure}

  \centering
  \centering \includegraphics [width=0.45\textwidth]{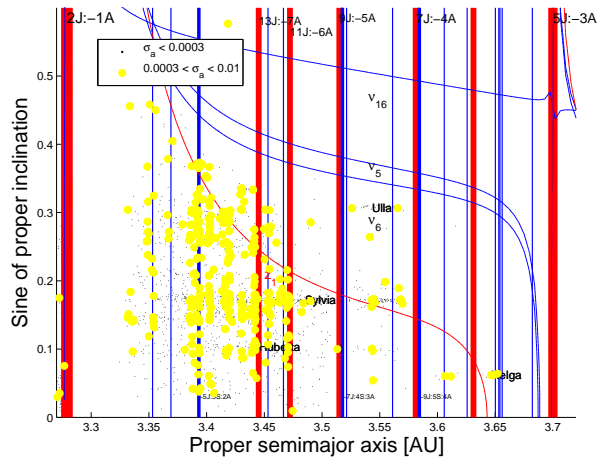}

\caption{An $(a,\sin{i})$ projection of proper elements for 
numbered and multi-opposition asteroids in the Cybele region.}
\label{fig: ai_da}
\end{figure}

A projection in the $(a,sin(i))$ plane of the proper elements (with their 
errors in proper $a$) for the surviving 1500 objects is shown in 
Fig.~\ref{fig: ai_da}.   Objects classified as
having unstable proper $a$ ($0.0003 < \Delta a < 0.01$~AU) are shown
as yellow full dots.  We then computed the location of two- and three-body
mean-motion resonances with the approach of Gallardo (2014).  This method
allows to compute the value of the semi-major axis of the resonance center
and to estimate the resonance strength through a parameter $R_S$ whose 
maximum value is one.  The location of three-body resonances was 
estimated up to order 30 including planets from Earth to Uranus, and
using the values of eccentricity, inclination and argument of pericenter
of 87 Sylvia, the asteroid with the largest family in the region.
Two-body resonances with $R_S$ up to $10^{-5}$ and three-body mean-motion 
resonances with $R_S$ up to $10^{-4}$ are shown 
as vertical red lines and blue lines, respectively (for 
simplicity we only identify three-body resonances with 
$R_S > 5 \cdot 10^{-2}$.  The thickness of
the lines is associated with the resonance strength.
Linear secular resonances are displayed as blue lines, while 
the inclined red line display the approximate location of the center
of the $z_1$ non-linear secular resonance (see Carruba (2010b) for
a description of the method used to plot the center of libration for
secular resonances).  The names of the asteroid families discussed
in Nesvorn\'{y} et al. (2015) and the Helga family of 
Vinogradova and Shor (2014) are also shown in the figure.

Most of the objects with unstable values of proper $a$ are perturbed
by two and three-body resonances such as the 13J:-7A, 11J:-6A, and
-5J:3S:2A.  As observed by several other authors (Milani and Nobili 
1985, Vokrouhlick\'{y} et al. 2010), known asteroid families tend to 
cluster in the regions with lower number density of resonances.
If we define highly inclined asteroids as those with
inclination higher than that of the center of the ${\nu}_6$ secular
resonance, only one numbered asteroid satisfies this criteria: 1373 
Cincinnati, with an inclination of $38.97^{\circ}$.  Generally speaking,
the ${\nu}_6$ secular resonance forms the upper boundary in inclination
of the Cybele region, while the 2J:-1A and 4J:-7A mean-motion resonances
with Jupiter delimit the region in semi-major axis for most of the
asteroids in the region, with the notable exception of the new (522) Helga
group, between the 4J:-7A and 5J:-3A mean-motion resonances, that could 
be the main belt family most distant from the Sun (Vinogradova and Shor 
2014), and is of particular interest because of the very peculiar
dynamics of 522 Helga, characterized by weak instabilities on extremely long
timescales, the so-called ``stable chaos'' of Milani and Nobili (1992, 1993).
Asteroids in the Cybele region do not generally experience planetary 
close encounters.   The region is also crossed by a web of non-linear
secular resonances, of which the most studied (Vokrouhlick\'{y} et
al. 2010) is the $z_1 = {\nu}_6 -{\nu}_{16} = g -g_6 +s-s_6$ secular
resonance.  We will further discuss secular dynamics in the region
in the next section.  


\section{Secular dynamics in the Cybele region}
\label{sec: sec_dyn}

Secular resonances occur when there is a commensurability between 
the proper frequency of precession of the argument of pericenter $g$
or of the longitude of the node $s$ of a given asteroid and a planet.
If the commensurability is just between one frequency of the asteroid
and one frequency of the planet, such as in the case of 
the ${\nu}_6 = g-g_6$ secular resonance, the resonance is linear.
Higher order resonances involving more complex combinations 
of asteroidal and planetary frequencies, such for instance
the $z_1= {\nu}_6 +{\nu}_{16} = g-g_6+s-s_6$ resonance, are called non-linear 
secular resonances.  Carruba (2009a) defines as likely resonators
the asteroids whose frequency combination is to within $\pm 0.3$ arcsec/yr
from the resonance center.  For the case of the $z_1$ resonance,
this means that $g+s = g_6+s_6 = 1.898$~arcsec/yr.  About 90\% of the
$z_1$ likely resonators in the Padua family area were found to be
actual librators, when the resonant argument of the resonance was analyzed.

\begin{figure*}

  \centering
  \begin{minipage}[c]{0.5\textwidth}
    \centering \includegraphics[width=3.5in]{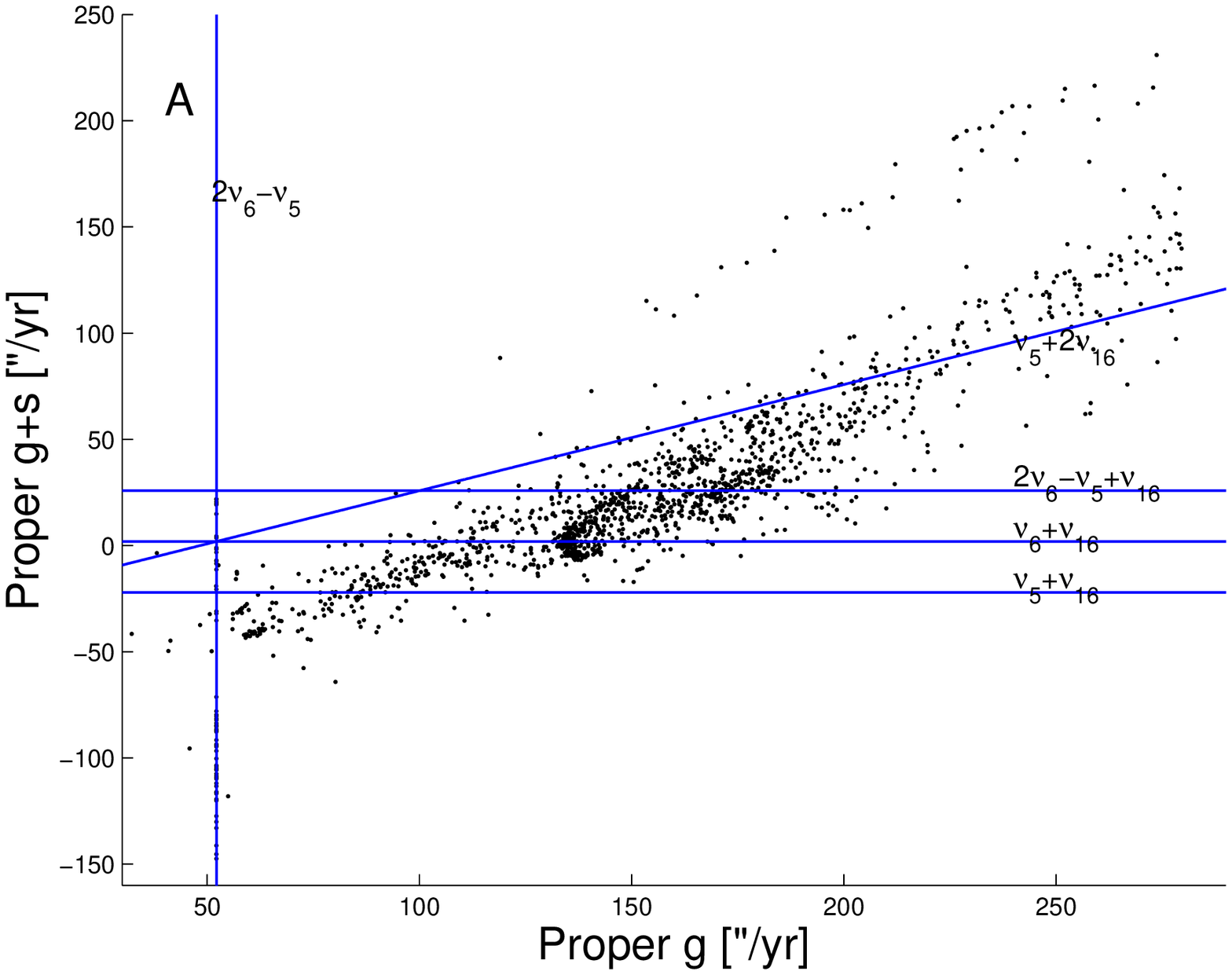}
  \end{minipage}%
  \begin{minipage}[c]{0.5\textwidth}
    \centering \includegraphics[width=3.5in]{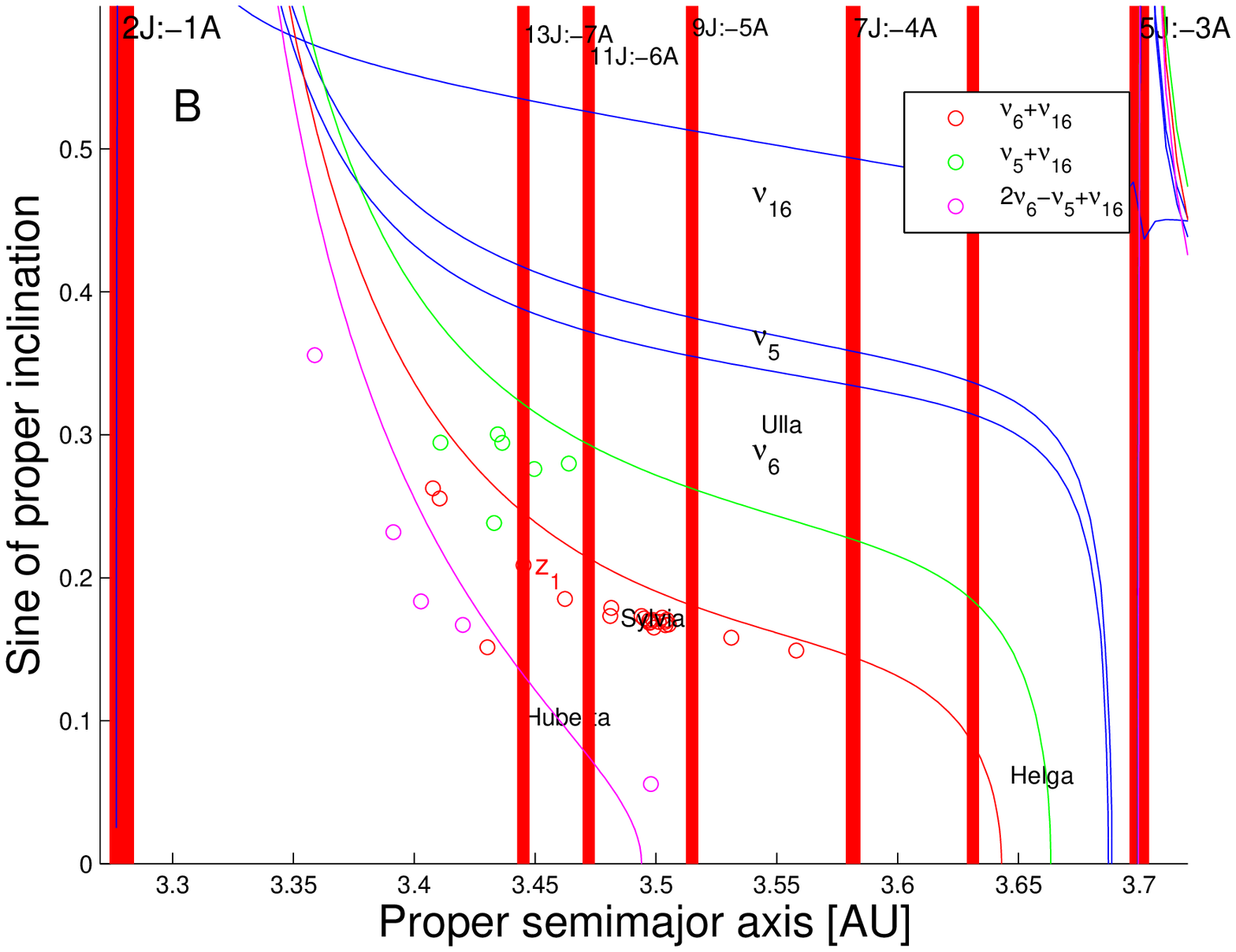}
  \end{minipage}

\caption{Panel A: a (g,g+s) projection of the 1485 asteroids with known proper
elements in our sample.  Blue lines identify the location of the secular
resonances listed in Table~\ref{table: sec_res} with a population
of ``likely resonators'' larger than 1.  Panel B: an (a,sin(I)) projection
of the actual resonators population listed in table~\ref{table: sec_res}.} 
\label{fig: cybele_ggs}
\end{figure*}

We selected likely resonators for all non-linear secular resonances
in the Cybele region up to order six following the procedure described in 
Carruba et al. (2013), and then verified that the appropriate
resonant argument for each of these asteroids was in a librating state.
Fig.~\ref{fig: cybele_ggs} displays a projection in the proper $(g,g+s)$
plane of asteroids in the region (panel A), as well as a projection
of asteroids found to be in librating states of non-linear secular
resonances in the region (panel B, only two-body resonances were reported
in this plot for simplicity).  Our results are summarized in 
Table~\ref{table: sec_res}, where we display the name of the resonances
with a population of asteroids larger than 1, the value of the combination
of planetary frequencies in the resonant argument, the number of likely
resonators, and the number of asteroids actually found in librating states.

No asteroids in the Cybele region were found to be in or near linear
secular resonances.  The large number of likely resonators found for
the $2{\nu}_6-{\nu}_5$ resonance is in fact an artifact caused by the
perturbation on the $g$ frequency value of objects near the separatrix
of the 2J:-1A mean-motion resonance (no asteroid was found to be 
in librating states of this resonance by our analysis).  
Three $g+s$ resonances have the largest
population of librators, the $z_1$ resonance with 22 asteroids (most of 
which are found in or near the Sylvia family (Vokrouhlick\'{y} et al. 2010)), 
the ${\nu}_5+{\nu}_{16}$ (six bodies), and the $2{\nu}_6-{\nu}_5+{\nu}_{16}$
with five objects.  Other local non-linear resonances with populations
of librators of just one object were not listed for the sake of brevity.

\begin{table}
\begin{center}
\caption{{\bf Main secular resonances in the Cybele region, frequency
    value, number of likely and actually resonant asteroids.}}
\label{table: sec_res}
\vspace{0.5cm}
\begin{tabular}{|c|c|c|c|}
\hline
                 &                   &                   &            \\
Resonance        & Frequency value   & Likely            & Actual     \\
argument         &   $[``/yr]$       & resonators        & resonators \\
                 &                   &                   &            \\
\hline
                 &                   &                   &            \\
                 &  g resonances     &                   &            \\
                 &                   &                   &            \\
$2{\nu}_6-{\nu}_5$&   52.229          &     44           &      0      \\
                 &                   &                   &            \\
                 & g+s resonances    &                   &            \\
                 &                   &                   &            \\
${\nu}_5+{\nu}_{16}$&  -22.088        &      6            &     6      \\
${\nu}_6+{\nu}_{16}$&  1.898          &     29            &    22      \\      
$2{\nu}_6-{\nu}_5+{\nu}_{16}$& 25.884 &      8            &     5      \\
                 &                   &                  &             \\
                 & g+2s resonances   &                  &             \\
                 &                   &                  &             \\
${\nu}_5+2{\nu}_{16}$& -48.433        &     3            &      0      \\
                 &                   &                  &             \\
\hline
\end{tabular}
\end{center}
\end{table}

Having revised the effect of local dynamics, we are now ready to analysis
the taxonomical properties of asteroids in the Cybele region.

\section{Compositional analysis: Taxonomy, SDSS-MOC4, and WISE data}
\label{sec: taxonomy}

In this section we will follow the same approach used by Carruba et al. 
(2014b) in their analysis of the Euphrosyne family.  A more in depth
description of the methods used can be found in that paper.
Only three objects had taxonomical data in three major photometric/spectroscopic
surveys (ECAS, SMASS, and S3OS2, see references in Carruba et al. 2014b): 
692 Hippodamia (S-type), 522 Helga, and 1373 Cincinnati (X-types). Using the 
classification method of De Meo and Carry (2013) that employs Sloan 
Digital Sky Survey-Moving Object Catalog data, fourth release (SDSS-MOC4 
hereafter, Ivezic et al. 2001)  to compute $gri$ slope and $z' -i'$ colors, 
we obtained a set of 238 observations of asteroids (including multiple 
observations of the same object) in the Cybele region.
This corresponds to 118 asteroids for which a SDSS-MOC4 
taxonomical classification and proper elements are both available. We 
found 44 X-types, 41 D-types, 23 C-types, 5 L- and S-types, and one A- and 
B-type object, respectively.
 
\begin{figure*}

  \centering
  \begin{minipage}[c]{0.5\textwidth}
    \centering \includegraphics[width=3.5in]{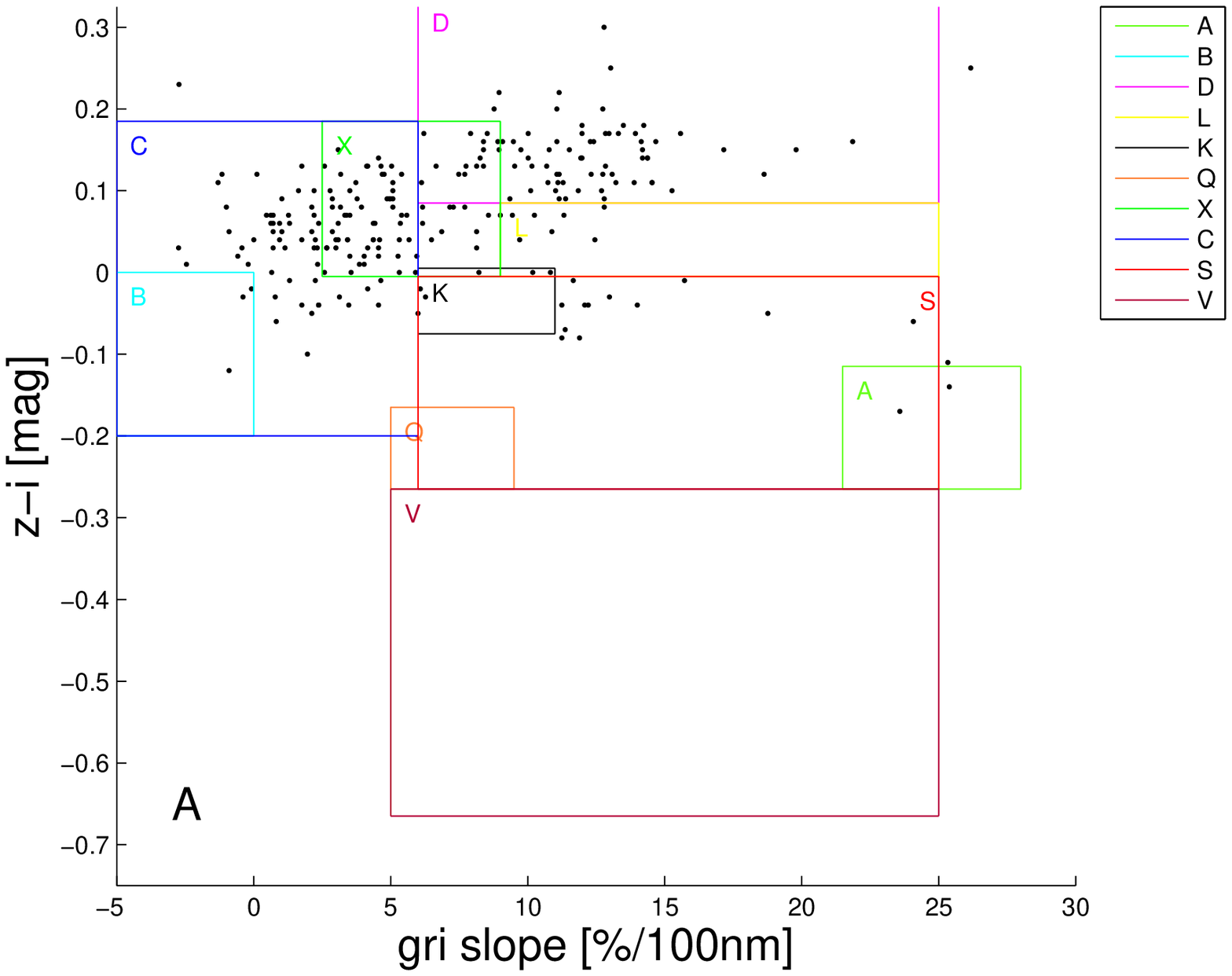}
  \end{minipage}%
  \begin{minipage}[c]{0.5\textwidth}
    \centering \includegraphics[width=3.5in]{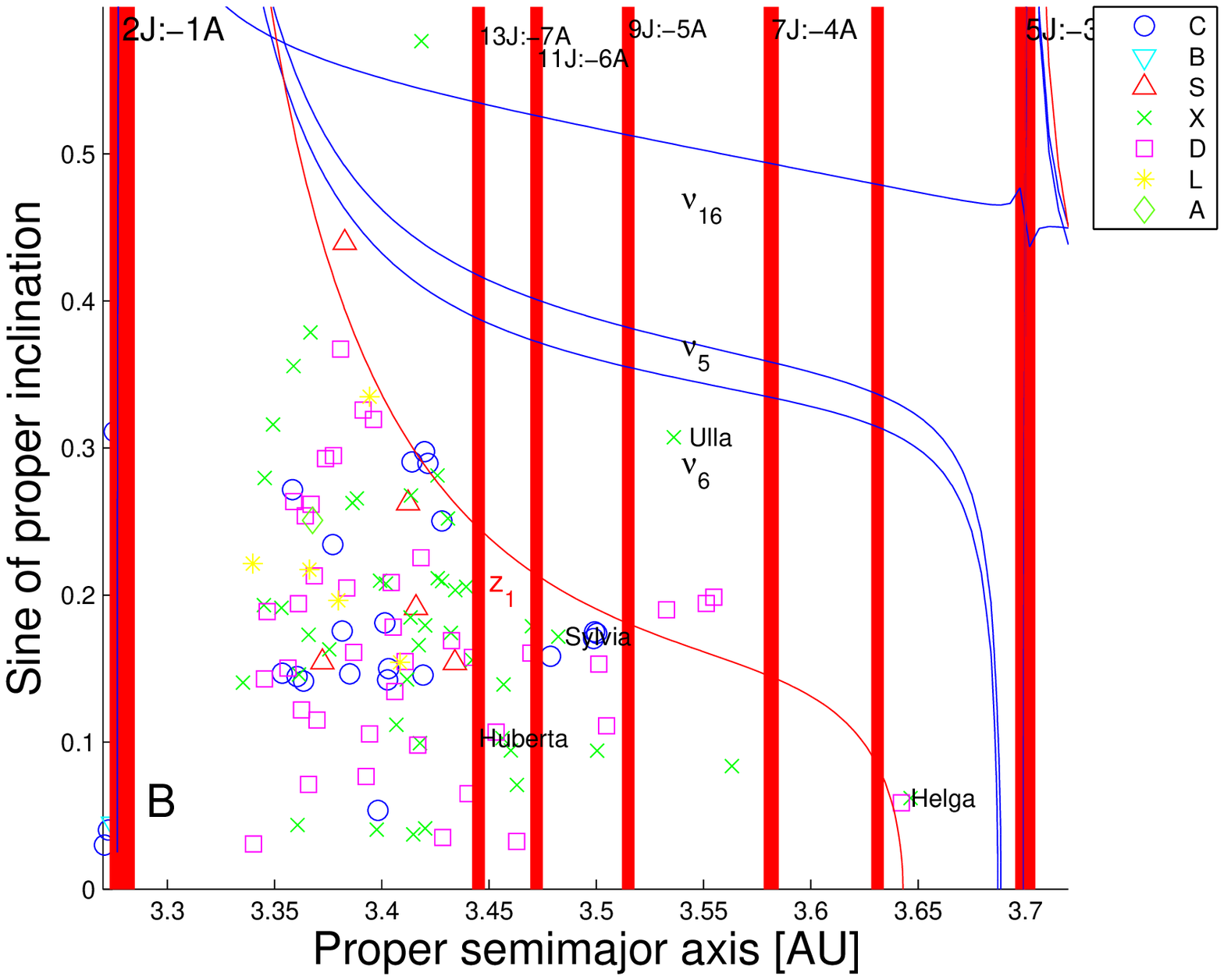}
  \end{minipage}

\caption{Panel A: a projection in the gri slope versus z-i plane of
the 238 observations in the SDSS-MOC4 catalogue for asteroids in the
Cybele region.  Panel B:  an $(a,sin(i)$ projection of asteroids
with taxonomical type in the same area.} 
\label{fig: cybele_sdss}
\end{figure*}

Fig.~\ref{fig: cybele_sdss}, panel A, displays a projection in the $gri$ slope 
versus $z-i$ plane for the 238 observations in the SDSS-MOC4 catalogue 
for asteroids in the Cybele region, while panel B shows an $(a,sin(i))$ 
projection of asteroids in the same region.  As in Carruba et al. (2013),
we found that the Cybele region is dominated by dark, primitive objects,
with a sizeable fraction of D-type bodies that were not identifiable
with the methods used in Carruba et al. (2013).  The Sylvia family
seems to be compatible with a CX-complex taxonomy.
The primitive, dark nature of most objects in the region is confirmed
by an analysis of the values of geometric albedos from the WISE
mission (Masiero et al. 2012).   We identified 568 asteroids with 
WISE albedo information in the Cybele region. Our analysis confirms
that of Carruba et al. (2013) (see Fig. 15): the vast majority of bodies
in the Cybele region has low albedos ($p_V < 0.15$), with only
11 asteroids with intermediate values ($0.15 < p_V < 0.3$), and
one single bright object with $p_V > 0.3$.   Using the values
of the diameters from WISE, when available (otherwise diameters 
are estimated using absolute magnitudes and the mean value 
of geometric albedo in the Cybele region, i.e., $p_V =0.067$ via 
Eq.~4 in Carruba et al. 2003), and the density of 87 Sylvia from Carry (2012),
we computed the masses of asteroids in the Cybele region, assumed
as homogeneous spheres.  For the few asteroids where an estimate
of the mass was reported in Carry (2012), we used the values 
from that paper.

\begin{figure}

\centering
\centering \includegraphics [width=0.45\textwidth]{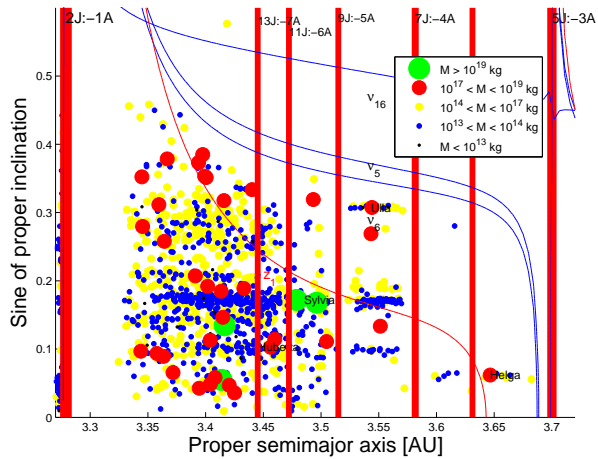}

\caption{An $(a,sin(i))$ projection of the asteroids in the Cybele region.  
The sizes of the dots are proportional to the asteroid masses, according
to the color code in the figure legend.}
\label{fig: mass_cybele}
\end{figure}

Fig.~\ref{fig: mass_cybele} shows an $(a,sin(i))$ projection of 
the family members, where the sizes of the dots are displayed according to 
the asteroid masses.  Of the four asteroids with masses larger than 
$10^{19}~kg$, 65 Cybele, 87 Sylvia, 107 Camilla, and 420 Bertholda,
only 87 Sylvia is currently associated with a dynamical family.  
Vokrouhlick\'{y} et al. (2010) suggested the possibility that 107 Camilla
and 121 Hermione, both of which are large binary asteroids, supposedly
generated during collisionary events, could have possessed family in the 
past that were dispersed through a combination of diffusion via the 
Yarkovsky effect and numerous weak resonances over timescales of a Gyr.  
We will further investigate this hypothesis later on in this paper.

Having now revised physical and taxonomical properties of 
local asteroids, in the next section we will start identifying 
dynamical groups in the region.

\section{Family identification}
\label{sec: fam_ide}

Several papers have discussed how to identify asteroid families using the 
hierarchical Clustering method (HCM).  Here we follow the approach 
of Carruba (2010b), see also Zappal\'{a} et al. (1995) for a discussion
the HCM.  Key parameters of the HCM are the nominal distance velocity
cutoff for which two nearby asteroids are considered to be related and 
possibly part of the same family, and the minimal number of bodies
for having a statistically significant family.  Using the approach of
Beaug\'{e} and Roig (2001) we found for our sample of 1500 asteroids
with proper elements in the Cybele region a nominal distance velocity 
cutoff of $138.2$~m/s, and a minimal family number of 25.

We constructed a stalactite diagram for asteroids in the region
using the approach of (Zappal\`{a} et al. 1990).  We selected (87) Sylvia,
the largest body of the most numerous family in the region, and used a 
value of the cutoff large enough that almost all Cybele 
asteroids were considered part of the family.  We then lowered the value 
of the cutoff and checked
if new dynamical groups appeared among the asteroids no longer connected
to the Sylvia family.

\begin{figure*}

  \centering
  \begin{minipage}[c]{0.45\textwidth}
    \centering \includegraphics[width=3.5in]{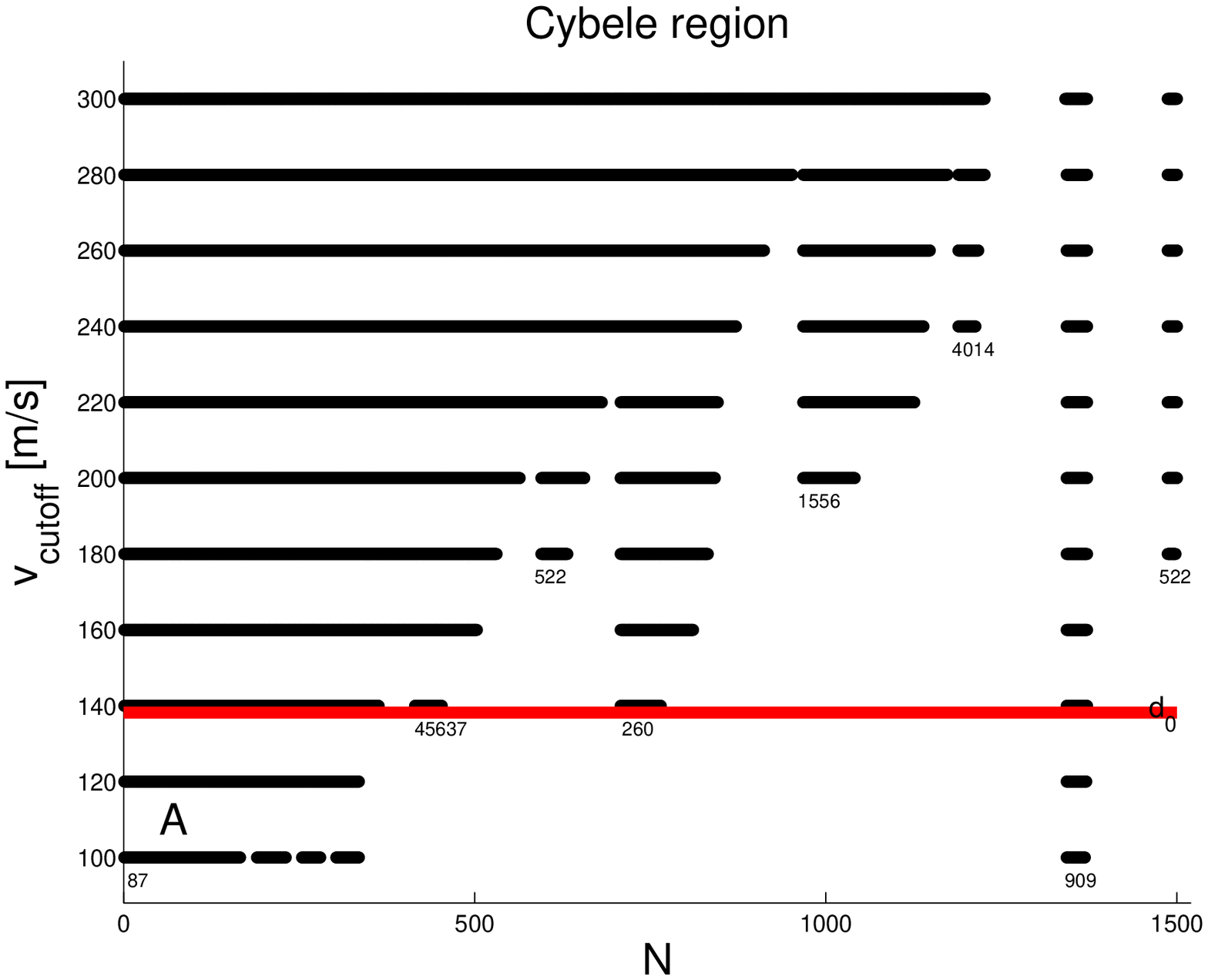}
  \end{minipage}%
  \begin{minipage}[c]{0.45\textwidth}
    \centering \includegraphics[width=3.5in]{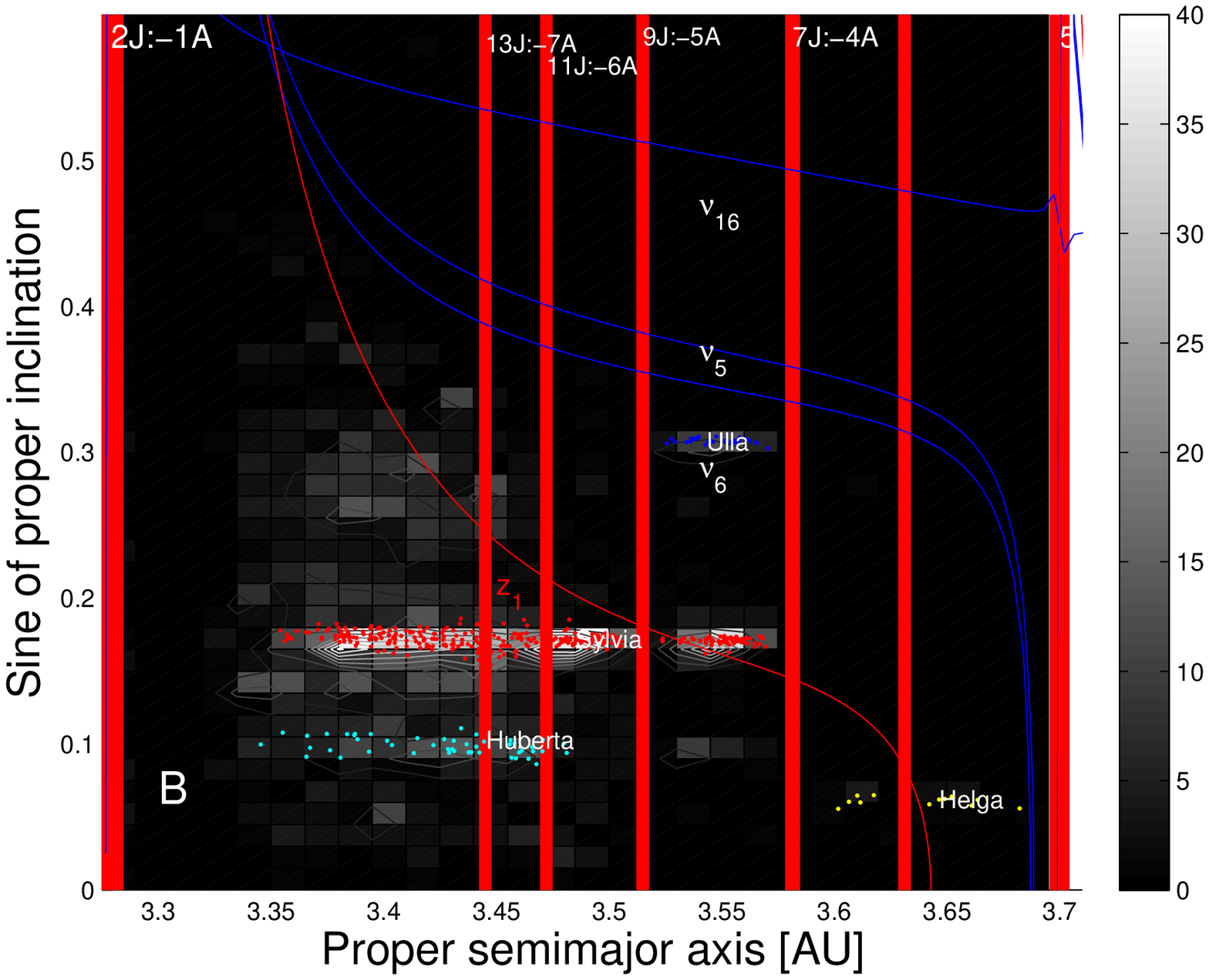}
  \end{minipage}

\caption{Panel A: stalactite diagram for the 1500 asteroids in the 
Cybele region.  Asteroids belonging to a family are identified by a 
black dots.  Panel B: contour plot of number density of asteroids
in the same region.  Colored dots show the location of members
of the four identified asteroid families.}
\label{fig: cybele_stal}
\end{figure*}

Fig.~\ref{fig: cybele_stal} displays our results.  Asteroids belonging 
to a family are identified by black dots, the horizontal red line 
shows the value of the nominal distance velocity cutoff as obtained
with the Beaug\'{e} and Roig (2001) approach.  At this value of nominal
cutoff only the three robust families of Nesvorn\'{y} et al. (2015) 
are detectable: 87 Sylvia (FIN 603, where FIN is the Family Identification
Number, as defined in Nesvorn\'{y} et al. 2015), 260 Huberta (just 
a candidate family in Nesvorn\'{y} et al. 2015), and 909 Ulla
(FIN 903).  The smaller family around 45637 2000 EW12, reported in 
Milani et al. (2014), disappears for smaller cutoffs and merges with the 
Sylvia family for higher cutoffs, and will not be considered as robust in 
this work. Of the new families proposed by Vinogradova and Shor (2014), we 
were able to only confirm the group around 522 Helga, and for the larger 
value of the cutoff of 180~m/s (this larger value is justified 
by the fact that these asteroids are found in a emptier region, with higher 
values of the mean distances among bodies with respect to the Cybele
region).  The proposed family of 643 Scheherezade, associated in our work
with asteroid 1556 Wingolfia, disappears for cutoffs of 200 m/s, and we
could not identify in this paper the proposed groups around 121 Hermione,
1028 Lydina, and 3141 Buchar.  We could also not positively identify
the ``rump'' families around 107 Camilla and 121 Hermione proposed
by Vokrouhlick\'{y} et al. (2010).  In that work the authors
suggested that the original families around these bodies, all
large asteroids with small satellites at large separations, 
configuration that could be the result of the formation of a collisional
family, could have possibly been lost because of interaction with 
the local dynamics in timescales of Gyr, and that, possibly, only
``rump'' families made by the largest remnants could have yet be found
near these two asteroids.  We did not identified any cluster 
near Hermione, and this asteroid merged with the Huberta family at 
a 200~m/s level.   There was a small clump of 16 asteroids 
near Camilla at 110~m/s, but apart from four asteroids and Camilla itself that 
had diameters larger than 5~km, the other objects are too small to  
have possibly survived 4 Gyr of dynamical evolution.   The possible existence
of rump families near these asteroids remains, therefore, yet to be proved, in 
our opinion.  Finally, the Sylvia family breaks down into 
four minor groups at cutoffs of 100 m/s or lower.\footnote{As in Carruba 
(2010b) we also searched for asteroid pairs in 
the region.  These objects, extremely close in proper element space, could
either be associated with double or multiple asteroid systems that recently 
separated, or be fragments launched by impacts onto very similar orbits 
(Pravec and Vokrouhlick\'{y} 2009).  The first two pairs with mutual distances 
less than 7.0~m/s were 113317 (2002 RL200) and 2007 RW290, and 216212 
(2006 UC73) and 239503 (2007 VH165) .}.    

\begin{table}
\begin{center}
\caption{{\bf Proper element based families in the Cybele region.}}
\label{table: Cybele_families}
\vspace{0.5cm}
\begin{tabular}{|c|c|c|c|c|c|}
\hline
     &     &    &    &    & \\
Name & $N$ & $N_{SDSS}$ & $N_{WISE}$ & Interlopers & Max. age [Myr]\\
     &     &    &    &    & \\
\hline
     &     &    &    &    & \\     
87 Sylvia   & 363 & 12& 99 & 8 & 3800\\
260 Huberta &  56 & 4 & 24 & 8 & 3000\\
522 Helga   &  12 & 2 &  1 & 0 &  800\\
909 Ulla    &  30 & 1 & 14 & 2 & 1200\\

    &     &    &    \\     
\hline
\end{tabular}
\end{center}
\end{table}

Our results are summarized in Table~\ref{table: Cybele_families},
that reports the smallest numbered family member, the number
of asteroids in the group, and those with data in the SDSS-MOC4
and WISE database.  There were 12 asteroids with SDSS colors in the Sylvia
family, 4 in the Huberta, 2 in the Helga, and one in the Ulla group.  All 
these objects are dark, primitive asteroids belonging to the C-, X-, 
and D-classes, and the WISE albedo data is also compatible with this scenario, 
since all family members with values of geometric albedos are dark 
bodies with $p_V < 0.15$. Fig.~\ref{fig: cybele_stal}, panel B, 
displays the orbital locations of our confirmed four families, shown as
colored dots, superimposed to contour plots of number density of asteroids
in the Cybele region.  We computed the number of asteroids per unit square 
in a 30 by 42 grid in the $(a,sin(i))$ plane, with $a$ 
between 3.26 and 3.70 and $sin(i)$ between 0.00 and 0.63.  Three inclination
regions of higher asteroidal number density can be identified in the map:
one around $\sin{(i)} \simeq 0.3$, associated with the Vinogradova and Shor
(2014) Scheherezade family, one at $\sin{(i)} \simeq 0.18$, occupied by the 
Sylvia family, and one at $\sin{(i)} \simeq 0.1$, associated with the Huberta
group.  The other minor local higher density regions are occupied 
by the Ulla and Helga families. 

As a next step in our analysis we obtain a preliminary estimate
of the age of the families and to eliminate possible dynamical interlopers
using the method of Yarkovsky isolines.  In this approach first the 
barycenter of the family is computed using the estimated diameters from WISE
(when available), and values of the density of the main family objects from
Carry (2012).  Then isolines of displacements caused by the Yarkovsky
effects for objects starting at the family barycenter are computed for
different estimated family ages, using values of the parameters
describing the Yarkovsky force from Bro\v{z} et al. (2013), Table 2.
For families older than 2000 Myr we also accounted for diffusion
caused by close encounters with massive asteroids, estimated to be of
the order of 0.02~AU over 4 Gyr.  Dynamical interlopers are objects
that reside beyond the maximum possible Yarkovsky isoline, and could
not have reached their current orbital position since the family formation.
A more in depth description of this method is available in Carruba et al. 
(2014b).

\begin{figure}

\centering
\centering \includegraphics [width=0.45\textwidth]{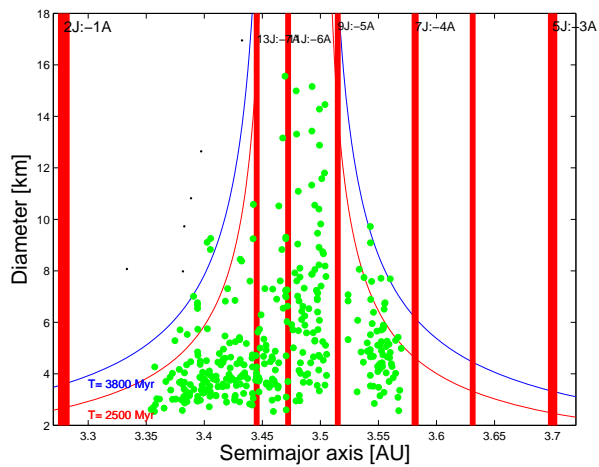}

\caption{A proper $a$ versus radius projection of 
members of the Sylvia family.  The red and blue curves
displays isolines of maximum displacement in $a$ caused by Yarkovsky
effect and close encounters with massive asteroids 
computed from the family barycenter after 2.5~Gyr (red line) 
and 3.8~Gyr (blue line).  Green full dots are associated with Sylvia family
members, black dots display the location of dynamical interlopers.}
\label{fig: iso_yarko}
\end{figure}

Fig.~\ref{fig: iso_yarko} displays the results of our method
for the Sylvia family, the largest group in the Cybele region.  We 
computed Yarkovsky (plus close encounters) isolines for 2.5 Gyr, the
estimated age of the family according to Vokrouhlick\'{y} et al. (2010),
and for 3.8 Gyr, the estimated age of the Late Heavy Bombardment (LHB)
according to Bro\v{z} et al. (2013).  Our results for this and other 
families in the area are summarized in Table~\ref{table: Cybele_families}, 
columns 5 and 6, where we report the number of dynamical interlopers 
and the estimated maximum age for each given family.  We found
8 dynamical interlopers (asteroids 107, 5914, 9552, 11440, 12003, 63050,
111526, 133402, 167625) in the Sylvia family, 8 (401, 3622, 4003, 
5362, 7710, 13890, 14330, 77837) in the Huberta group, and 2 (60042,
114552) in the Ulla family.   The Sylvia family could be 
one of the few groups created during the LHB, and the Huberta family
is characterized by an asymmetry between lower and higher semi-major
axis asteroids, with a larger fraction of the family closer to the Sun (the
estimate of the age significantly decreases if we disconsider a larger
fraction of the members closer to the Sun).
The Ulla and Helga groups have relatively small populations of members
(29 and 12, respectively), which undermines the possibility of obtaining 
more precise estimates of these groups ages.  In the next section
we will use the so-called Yarko-Yorp Monte Carlo method of Vokrouhlick\'{y}
et al. (2006a, b, c), modified to account for new developments
in our understanding of the YORP effect, to try to refine
the preliminary estimates of the family ages obtained in this section.

\section{Chronology}
\label{sec: chron}

\begin{figure*}

  \centering
  \begin{minipage}[c]{0.45\textwidth}
    \centering \includegraphics[width=3.5in]{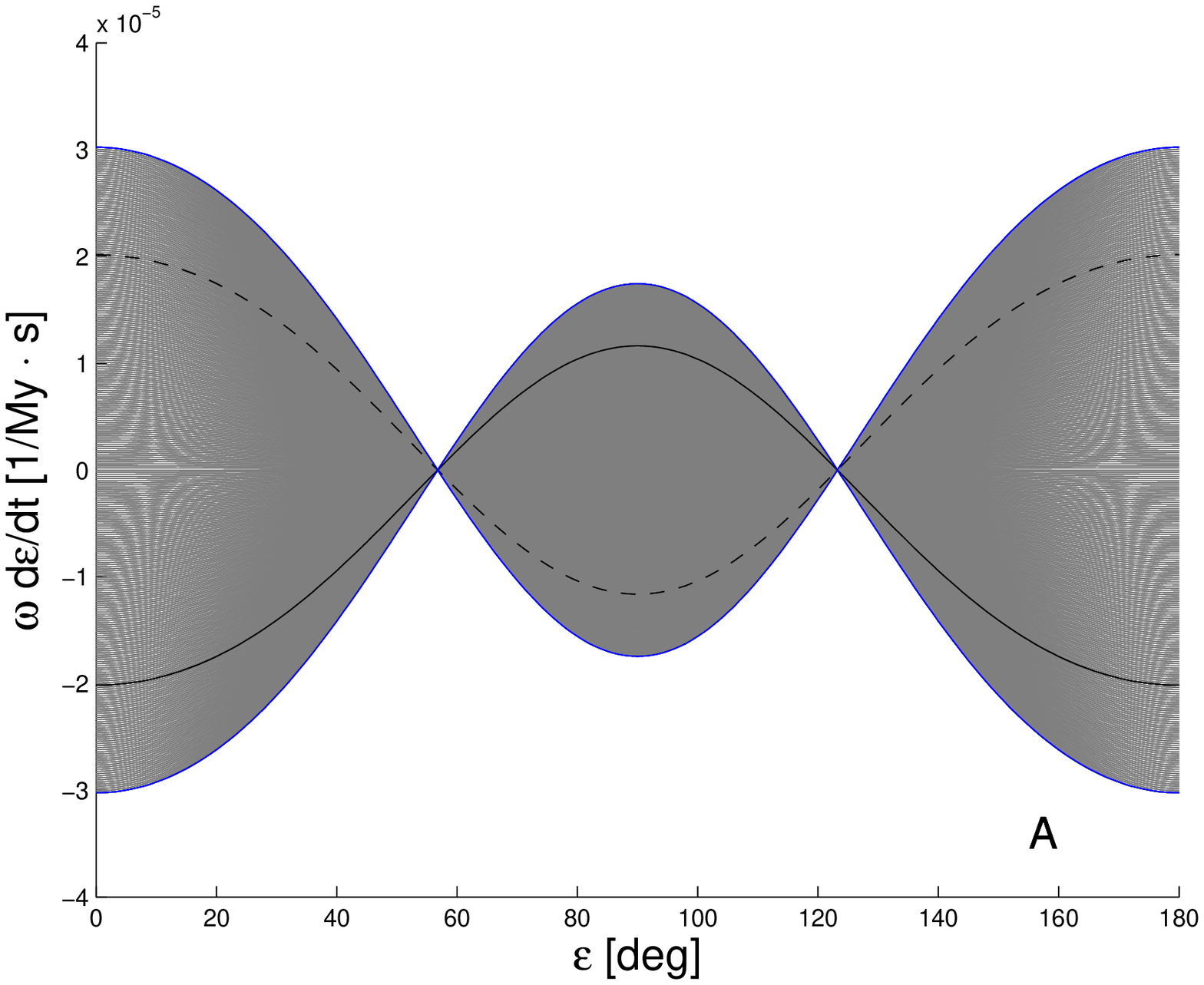}
  \end{minipage}%
  \begin{minipage}[c]{0.45\textwidth}
    \centering \includegraphics[width=3.5in]{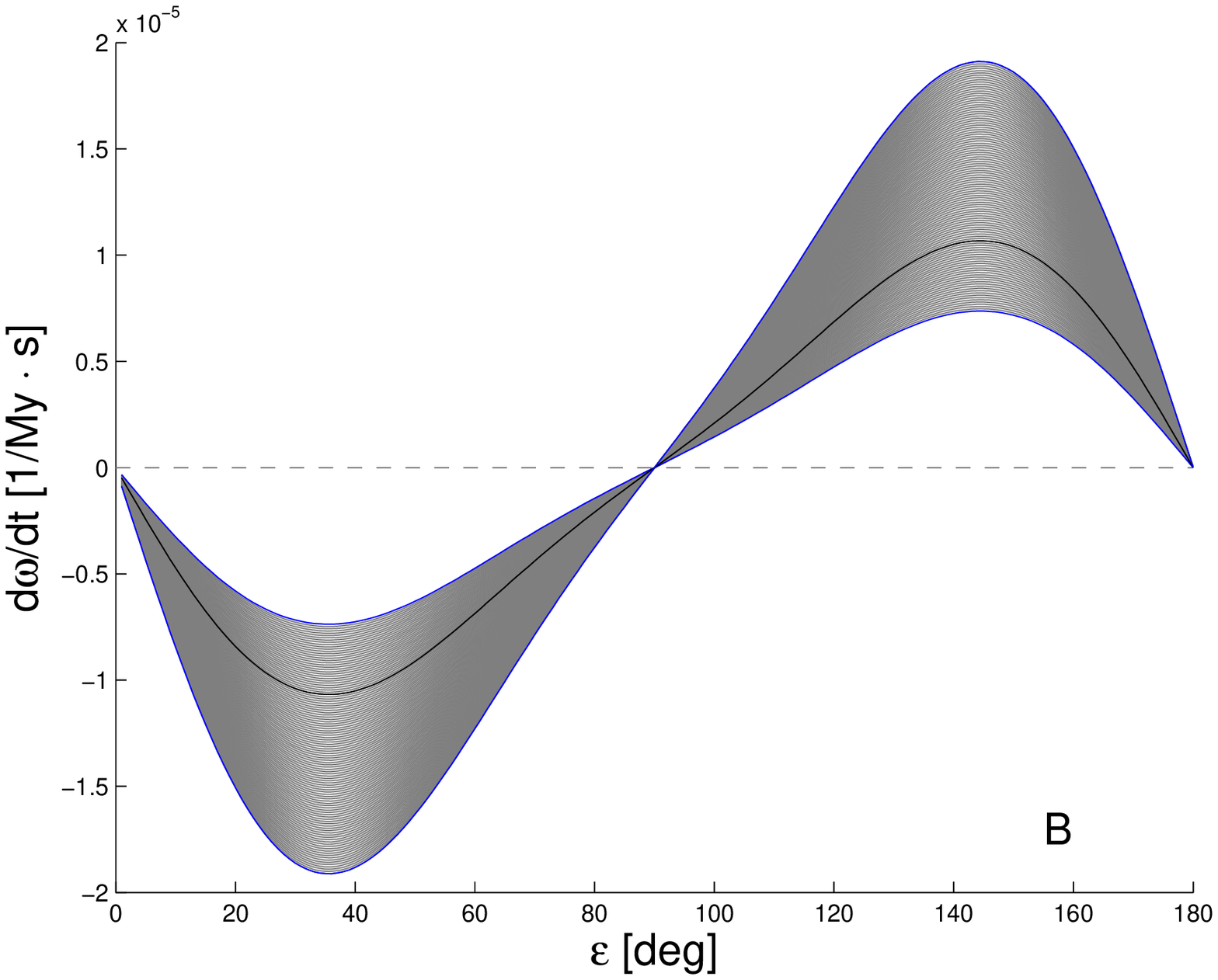}
  \end{minipage}

\caption{Dependence of the $g-$ (panel A) and $f-$ (panel B) 
functions on the spin obliquity $\epsilon$. 
Panel A displays $g = \omega (d \epsilon /dt)$, while in panel B we show
$f = d\omega /dt$.   Blue lines display the curves of maximum
variance of the $f-$ and $g-$ functions, while the gray zone is
associated with the variance of the results.  The full and dotted 
lines show the behavior of solution for accelerating and 
decelerating bodies, respectively.} 
\label{fig: fg_func}
\end{figure*}

Monte Carlo approach to obtain estimates of the family age and ejection 
velocity parameters were pioneered by Vokrouhlick\'{y} et al. (2006a, b, c) 
for the Eos and other asteroid groups.  Here we modified that method to 
also account for the stochastic version of the YORP effect, as modeled 
by Bottke et al. (2015).  To account for the effect of shape changes on 
the torques acting on a given asteroid during YORP cycles, recently Bottke 
et al. (2015) introduced the so-called 
``stochastic YORP'' model.  In this model, a different torque 
solution is chosen every time the ${\tau}_{YORP}$ timescale (of the order 
of 1 Myr) is exceeded.  This approach changes the evolution of 
the asteroid periods with respect to the ``static YORP'' effect in which 
shapes remained fixed between YORP cycles, 
but does not affect the time-behavior of the asteroid obliquities, which are 
still driven toward the end states of $0^{\circ}$ and $180^{\circ}$. 
To estimate the age and other parameters of the given asteroid family, 
basically, first the $(a,H)$ values of asteroids are mapped into 
a $C$-parameter using the relationship:

\begin{equation}
0.2H=log_{10}(\Delta a/C).
\label{eq: target_funct_C}
\end{equation}  

\noindent
Then, fictitious  distributions of asteroids for different values of 
the ejection velocity (and other) parameters are evolved under the 
influence of the Yarkovsky effect, both diurnal and seasonal versions, 
the YORP effect, and, in some cases, the long-term effect of close encounters 
with massive asteroids (Carruba et al. 2014a).  
Changes in the angular rotation velocity $\omega$ and spin obliquity (the
inclination of the spin axis with respect to the normal to the orbital
plane of the asteroid) $\epsilon$ caused by the YORP effect are computed 
by solving the system of differential equations:

\begin{eqnarray}
\omega \frac{d \epsilon} {dt} & = & g(\epsilon)\\ 
\frac{d \omega} {dt} & = & f(\epsilon),
\end{eqnarray}

\noindent where the $g-$ and $f-$ functions are displayed in 
Fig.~\ref{fig: fg_func} (see also Bottke et al. 2015).   These functions 
were computed for a $D = 2~$km body at $a= 2.5$~AU, with a bulk density of
$\rho_{bulk} = 2500~kg/m^3$ and a period of 8~hrs.  The strength 
of the YORP effect is proportional to $1/(\rho_{bulk} a^2 D^2)$.
The gray zone in the figure in between the blue lines displays 
the variance of the results for various asteroidal shapes in 
\v{C}apek and Vokrouhlick\'{y} (2004).  The full and dotted 
lines show the behavior of solution for accelerating and 
decelerating bodies, respectively.  While in the so-called static YORP
the $f-$ and $g-$ functions are constant between end states of the YORP
cycle, in the stochastic YORP, to account for the changes in the shape
of the asteroid produced by increasing or decreasing spin periods,
these two functions are randomly changed on time-scale of 1~Myr (see
Bottke et al. 2015 for more details on the modeling of this effect).
Once a $C$-target values distribution has been obtained for test 
particles subjected to the Yarkovsky, YORP, and, possibly, other force,
the real and fictitious distributions of $C$-target values are then 
compared using a ${\chi}^2$-like variable ${\psi}_{\Delta C}$  
(Vokrouhlick\'{y} et al. 2006a, b, c), whose 
minimum value is associated with the best-fitted solution.

Of the four families identified in this work, only the Sylvia and (barely) the
Huberta group have enough members to have a distribution of $C$-target values 
large enough for the method to be appliable.  We will start our analysis
by investigating the Sylvia group.

\subsection{Sylvia family}
\label{sec: chron_sylvia}

As discussed in Sect.~\ref{sec: fam_ide}, the Sylvia family may be as old 
as 4.2~Gyr.  Since in the past the solar luminosity was weaker, and 
this affects the strength of the Yarkovsky force, following the approach 
of Vohkroulick\'{y} et al. (2006a) we modified the Yarko-Yorp code
to also account for older values of the solar luminosity $L(t)$ given by 
the relationship:

\begin{equation}
L(t)=L_{0} \left[1+0.3 \left( 1-\frac{t}{t_0}\right)\right]^{-1},
\label{eq: sol_const}
\end{equation}

\noindent
where $L_{0}$ is the current value of the solar constant
($1.368 \cdot 10^3 W/m^2$), $t_0 \simeq 4.57~Gyr$ is the age of the Sun,
and $t$ is the time measured from the Solar System formation.  
We then computed values of the ${\chi}^2$-like parameter for 28
values of the ejection velocity parameter $V_{EJ}$ up to $140$~m/s (the 
current escape velocity from 87 Sylvia is 111.74~m/s, values of $V_{EJ}$  
higher than the escape velocity were not likely to occur, according to 
Bottke et al. 2015), using
values of the Yarkovsky parameters (and other parameters needed
by the Yarko-Yorp model such as $C_{Yorp}$, $c_{reorient}$ and ${\delta}_{YORP}$
whose descriptions and used values can be found in Bottke et al. 2015), 
already discussed in previous sections and typical of C-complex families
as Sylvia (see also Table 2 of Bro\v{z} et al. 2013).

As in previous papers, to avoid problems caused by possible fluctuations 
in the $N(C)$ distribution, we also computed an average $C$ distribution 
for the family obtained as a mean of $C$ distributions, each corresponding
to five values of $a_c$ (in the range $[3.4785,3.4789]~AU$, 
near the family barycenter).  To avoid the problem
of small divisors, we did not considered 3 intervals in $C$ with less 
than 3 asteroids (about 1\% of the total population of 354), leaving 
us with a total of 25 intervals in $C$.

\begin{figure}
  \centering
  \centering \includegraphics [width=0.45\textwidth]{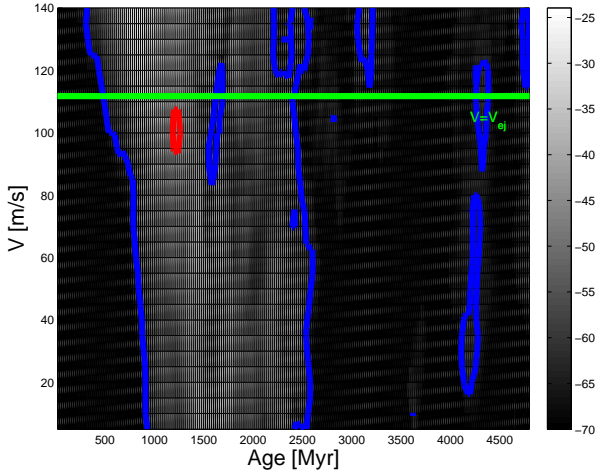}

\caption{Target function ${\psi}_{\Delta C}$ values in the ($Age,V_{EJ}$) plane 
for Sylvia families simulated with the stochastic Yarko-YORP Monte Carlo model,
modified to account for lower solar luminosity in the past.
The red and blue lines identify the values of the target 
function associated with the two given probabilities of the 
simulated distribution agreeing with the real one.  The horizontal green 
line shows the value of the ejection velocity 
from 87 Sylvia.}
\label{fig: cont_sylvia}
\end{figure}

Fig.~\ref{fig: cont_sylvia} displays our results. As discussed in 
Vokrouhlick\'{y} et al. (2006a), the estimated age of the family is 
about 10\% older when the effect of lower past solar luminosities is 
accounted for.  For values of $V_{EJ}$ lower than the escape velocity 
from 87 Sylvia, we found two minima, one (red curve in 
Fig.~\ref{fig: cont_sylvia}) for $T = 1220^{+40}_{-40}~Myr$ 
and $V_{EJ} = 95^{+13}_{-3}$~m/s, for ${\psi}_{\Delta C} = 24.5$, corresponding 
at a confidence level of 56.7\% of the two distribution being compatible
(Press et al. 2001), and a much weaker one (blue curve) for 
$T = 4220^{+100}_{-120}~Myr$ and $V_{EJ} = 35^{+45}_{-15}$~m/s, 
for ${\psi}_{\Delta C} = 58.5$, corresponding at a confidence level of just 0.8\%.
The possible existence of two distinct minima may either suggests that 
i) the Sylvia family could be the result of two distinct 
cratering events, with the less likely solution corresponding to an older
cratering event, or ii) the secondary solution is just a statistical fluke.  
We will further investigate these two scenarios in the next section.

\subsection{Huberta family}
\label{sec: chron_huberta}

\begin{figure}
  \centering
  \centering \includegraphics [width=0.45\textwidth]{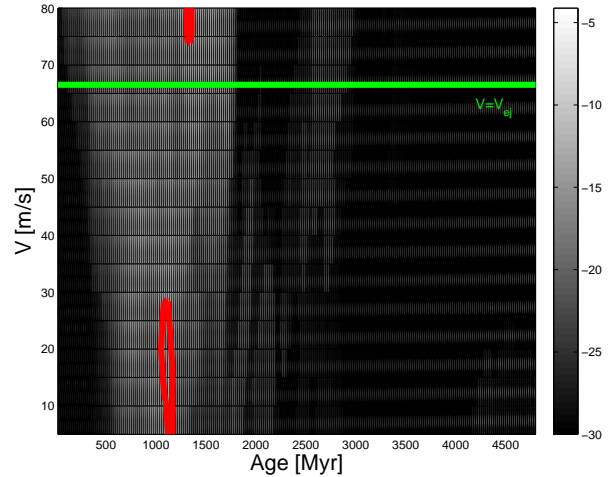}

\caption{Target function ${\psi}_{\Delta C}$ values in the ($Age,V_{EJ}$) plane 
for Huberta families simulated with the same approach used for the Sylvia
family.}
\label{fig: cont_huberta}
\end{figure}

After eliminating the dynamical interloper, the Huberta family still possessed
48 members.  We created a $C$-target function distribution with 9 intervals
in $C$ equally spaced by $\Delta C = 20 \cdot 10^{-6}$ and starting at 
$C = -9 \cdot 10^{-5}$. Again, we used five values of $a_c$ near the
family barycenter to compute the distribution of $N_{obs}(C)$, that was
then averaged among the five values. The escape velocity from 260 Huberta 
is $66.5$~m/s (displayed as a horizontal green line in 
Fig.~\ref{fig: cont_huberta}), we expect therefore that the 
best-fitted solution should have values of $V_{EJ}$ lower than that.

Fig.~\ref{fig: cont_huberta} displays our results:  we found a minimum
for $T=1100 \pm 50~Myr$ and $V_{EJ} = 15^{+13}_{-15}$~m/s, at a confidence
level of 91.8\%.  We did not identify significant secondary minima for the 
Huberta family. The dynamical evolution of this and other families in 
the Cybele region will be further investigated in the next section.

\section{Dynamical evolution of families in the Cybele region}
\label{sec: dyn_cybele}

\subsection{Future dynamical evolution}
\label{sec: fut_dyn}

To further investigate the dynamical evolution of dynamical groups in the
Cybele region, we performed numerical integration with the 
$SWIFT-RMVSY$ integrator of Bro\v{z} (1999), able 
to model the diurnal and seasonal Yarkovsky force and the effect of 
close encounters of massless particles with massive planets, 
and with the new $SYSYCE$ integrator (Swift$+$Yarkovsky$+$Stochastic
Yorp$+$Close encounters), that modifies the S\~{a}o Paulo integrator of 
Carruba et al. (2007), a code that modeled the Yarkovsky force and 
close encounters of massless particles with massive bodies, to also 
include the stochastic YORP effect, as described in Bottke et al. (2015).  
For completeness, we also modified $SWIFT$-$RMVSY$, so as to account for 
both the static and stochastic versions of the YORP effect.  While we do not
consider the long-term effect of close encounters with massive asteroids
in this work, the $SYSYCE$ integrator can in principle also account for 
this perturbation.

\begin{figure}

  \centering
  \centering \includegraphics [width=0.45\textwidth]{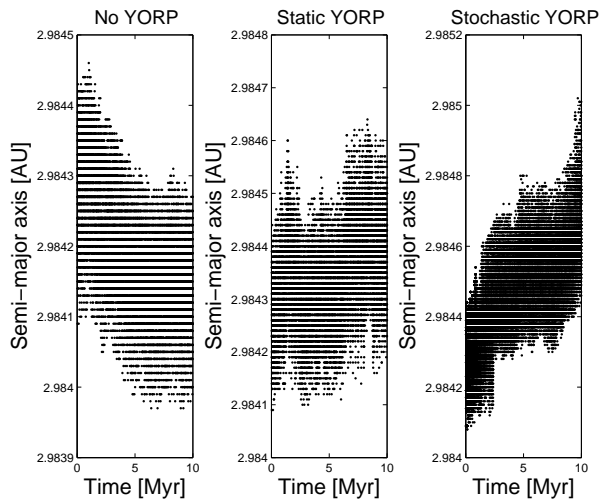}

\caption{Time evolution of the osculating semi-major axis of a particle
integrated under the effect of the Yarkovsky effect and no YORP force (left
panel), under the static YORP effect (central panel), and under stochastic
YORP (right panel).}
\label{fig: symplectic_yorp}
\end{figure}

Fig.~\ref{fig: symplectic_yorp} displays the time evolution of the
osculating semi-major axis of a test particle in the Euphrosyne family
region integrated over 10 Myr under the gravitational effect of all
planets and (31) Euphrosyne\footnote{The mass of (31) Euphrosyne 
($M = 1.27 \pm 0.65 \cdot 10^{19}~kg$) was taken from Carry (2012).}, with
the S\~{a}o Paulo integrator (No YORP, left panel); using the new
integrator including the static version of the YORP effect (Static
YORP, central panel); and using the new integrator while including the
stochastic YORP effect (Stochastic YORP, right panel).  For our runs,
we used the optimal values of the Yarkovsky parameters discussed in 
Bro\v{z} et al. (2013) for C-type asteroids.  The initial spin obliquity 
was $0^{\circ}$ for the simulation without the YORP effect and random 
in the other two cases.
Normal reorientation timescales due to possible collisions as
described in Bro\v{z} (1999) were considered for all runs.  As can be
seen in the figure, the particle experienced just a single event of
reorientation at about 1 Myr, and then evolved under the Yarkovsky
force with a retrograde obliquity, toward smaller values of semi-major
axis.  In the simulation with the static YORP integration,
reorientations occurred twice, once because of a collision (at about 1
Myr), and once because the period of the test particle reached a limit
value (2 or 1000 hrs, see Bottke et al. 2015 for a discussion
of what happens at the YORP cycle end-states), at about 3
Myr.  As a consequence, the test particle started evolving toward
larger values of $a$, then reoriented its spin axis and evolved toward
smaller $a$ values, and finally reoriented once again and evolved
toward larger values of $a$.  More interesting was the case of the
stochastic YORP integration. Here the shape of the test particle 
(but not its orientation) was changed every Myr.  As a consequence,
the test particle underwent a less severe random walk, tending to
generally increase its semi-major axis over the integration.
Similar behavior was observed for several other test runs performed
with the new symplectic code.  Results for simulations with the
modified versions of the $SWIFT$-$RMVSY$ integrator are compatible,
within numerical errors, with that of the SYSYCE integrators, for both
the static and stochastic version of the YORP force.

Having tested the new integrator, for our runs with Cybele asteroids again we 
used the optimal values of the Yarkovsky parameters discussed in Bro\v{z} 
et al. (2013).  Normal reorientations timescales due to possible collisions, 
as described in Bro\v{z} (1999), were considered for all runs with the 
$SYSYCE$ integrator, but not for those with $SWIFT-RMVSY$, so as to obtain 
the maximum possible drift rates for these simulations.  
All particles were subjected to the gravitational influence of all planets.
For the simulations with $SWIFT-RMVSY$ we used for each family member, 
as determined in Sect.~\ref{sec: fam_ide}, two sets of orbital obliquity,
$0^{\circ}$ and $180^{\circ}$, so as to maximize the strength of the Yarkovsky 
force.  Random spin obliquities were assigned to the test particles
for the simulations with the $SYSYCE$ integrator.  We integrated our
test particles over 4.0 Gyr, under the influence of all planets.

\begin{table}
\begin{center}
\caption{{\bf Time needed for the family to have less than 8 members
(dispersion time) for simulations with the $SWIFT-RMVSY$ and $SYSYCE$ 
integrators.}}
\label{table: dispersion_times}
\vspace{0.5cm}
\begin{tabular}{|c|c|c|}
\hline
       &                     &                 \\
Family & Dispersion time     & Dispersion time \\
       & $SWIFT-RMVSY$ [Myr] & $SYSYCE$ [Myr]  \\
\hline
       &                     &                 \\
Sylvia & 2650                & 4020            \\
Huberta& 1650                & 2270            \\
Ulla   & 650                 &  705            \\
Helga  & 305                 &  320            \\
       &                     &                 \\
\hline
\end{tabular}
\end{center}
\end{table}

\begin{figure}

  \centering
  \centering \includegraphics [width=0.45\textwidth]{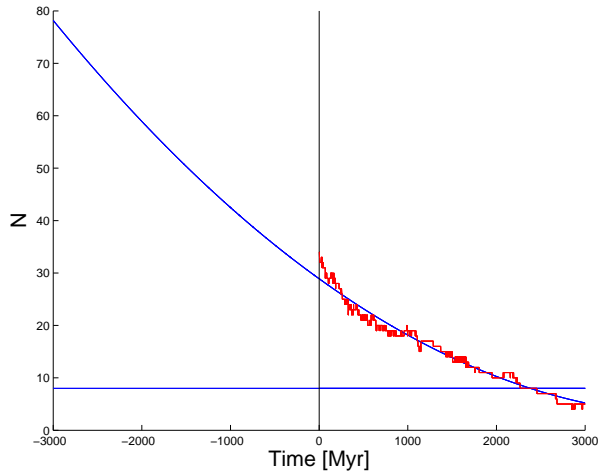}

\caption{Number of simulated asteroids in the Huberta family region as 
a function of time (red line).  The blue line displays results of a best-fitted
second order polynomial to the $(time,N)$ function extended to the maximum 
possible age of the Huberta family (3 Gyr).  The vertical black line displays
present time, and the horizontal blue line the minimum number of asteroid
needed to recognize the family as a clump (8).}
\label{fig: huberta_sysyce}
\end{figure}

We first obtained synthetic proper elements for our test particles with 
the methods described in Sect.~\ref{sec: prop_el} and computed the number of 
particles that remained in the four studied family regions, defined as boxes 
in the $(a,e,sin(i))$ domain delimited by the maximum and minimum value of each 
proper element.  Fig.~\ref{fig: huberta_sysyce} displays the number of
asteroids, simulated with the SYSYCE integrator, in the Huberta family 
region as a function of time (red line).   We best-fitted a second-order
polynomial to these data and extrapolated our results to 3 Gyr ago, the
maximum possible age of the Huberta family as obtained with the 
method of the Yarkovsky isolines.   The number of family members in that 
region will become less than 8, the minimum number of objects to identify
a dynamical group, in 2.27 Gyr.  We extrapolated that 3.00 Gyr ago the family 
may have had 80 members, assuming that the rate of population loss
obtained from our simulations remained constant.

Our results are also summarized in Table~\ref{table: dispersion_times}, where
for each family we show the time needed to have only eight members in the 
family region (dispersion time) for simulations with the $SWIFT-RMVSY$ and 
$SYSYCE$ integrators.  As expected dispersions times obtained also considering
the stochastic YORP effect tend to be higher than analogous times obtained
with the simpler $SWIFT-RMVSY$ integrations, where no reorientations were 
considered (but timescales should be lower when compared with the dispersion 
times obtained with the static YORP effect, since the static YORP effect
tends to change the shape, but not the orientation of the spin
axis, Bottke et al. 2015).  The dispersion time for the simulation with 
the stochastic YORP effect for the Sylvia family was slightly longer than 
the integration length and was computed extrapolating our results 
with the method described for the case of the Huberta family.

\subsection{Past dynamical evolution}
\label{sec: past_dyn}

We then turned our attention to the past evolution of our studied families.
For the two families for which it was possible to apply our Yarko-Yorp approach 
of Sect.~\ref{sec: chron}, i.e., Sylvia and Huberta, we generated fictitious
families with the ejection parameter $V_{EJ}$ previously found and the number
of objects the family that they were supposed to have at the time of their 
formation.  For the Huberta family, we extrapolated this number to the 
age estimated with the Yarko-Yorp method of Sect.~\ref{sec: chron}, 
i.e. 1100 Myr, and for the maximum possible age obtained with the method 
of the Yarkovsky isolines, i.e., 3000 Myr.  We obtained two families, 
with 48 and 80 members each.  For the Sylvia family we extrapolated the 
number of objects to the time at which the family was formed according to 
the results of Sect.~\ref{sec: chron}, i.e., 1220 Myr and $\simeq$ 1000 members.
Considering that information about the initial number of members of the 
possible oldest Sylvia family discussed in Sect.~\ref{sec: chron} is not 
easily available, we then also integrated the same family of 1000 members
of the first Sylvia integration over the possible age of the first family,
i.e., 4220 Myr.  Since Bottke et al. (2012) suggested that the Late
Heavy Bombardment may have started earlier than previously thought,
with estimates as early as 4.1 Gyr ago (against the 3.8 Gyr previously
assumed), this second age estimate of the Sylvia may possibly be in 
agreement with our new understanding of the LHB time of occurrence.
As done in Sect.~\ref{sec: chron} for the Sylvia family, 
we also modified the $SYSYCE$ integrator to account for the lower past 
solar luminosity, according to Eq.~\ref{eq: sol_const}.  Finally, 
in view of the fact that Vokrouhlick\'{y} et al. (2010) suggested 
that families around 107 Camilla and 121 Hermione could have formed
in the past and then dispersed, we created fictitious families around
these objects of 80 members each (the same estimated initial number of 
members of the Huberta family), and integrated these objects since
3800 Myr ago, i.e., after the most recent estimate for the 
end of the Late Heavy Bombardment (LHB).
Because of the very small number of members of the Helga and Ulla group, 
we did not integrated these groups in the past, but in first approximation
we can assume that dispersion times in the past for these groups should 
be of the same order of those obtained with integrations into the future.

\begin{table}
\begin{center}
\caption{{\bf Dispersion times for results of past dynamical evolution
of families in the region.}}
\label{table: dispersion_times_past}
\vspace{0.5cm}
\begin{tabular}{|c|c|c|}
\hline
       &                     &                 \\
Family & Initial time        & Dispersion time \\
       &   [Myr]             &    [Myr]        \\
\hline
       &                     &                 \\
Sylvia & 4200                & $ >4200$        \\
Sylvia & 1250                & $ >1250$        \\
Huberta& 3000                & $1450\pm40$     \\
Huberta& 1100                &  $920\pm50$     \\
Hermione&3800                &  $930\pm10$     \\
Camilla &3800                & $1460\pm50$     \\
       &                     &                 \\
\hline
\end{tabular}
\end{center}
\end{table}

Table~\ref{table: dispersion_times_past} displays the dispersion times
for our integrated families (third column), using the same box-like
criteria discussed in Sect.~\ref{sec: fut_dyn}.  For the simulation
with the Sylvia family over 1250 Myr, we found that the family 
was still detectable at the end of the run, with 104 objects in the
Sylvia family region, thus confirming our previous estimate of 
Sect.~\ref{sec: chron_sylvia}.  Extrapolating the current loss rates to 
the past, we estimated that the Sylvia family may have been $\simeq 7$ 
time more numerous at the time of its formation.  At the end of the
run 36.0\% of the asteroids with $10 < D < 12$~km, 30.2\% of those
with $8 < D < 10$~km, and 18.5\% of those with  $6 < D < 8$~km were
still in the Sylvia family region, so suggesting that only the 
largest fragments of an hypothetical LHB Sylvia family may still 
be visible today in the Cybele region.   Of more interest were the 
results of the Sylvia simulations starting 4200 Myr ago.
  
\begin{figure}

  \centering
  \centering \includegraphics [width=0.45\textwidth]{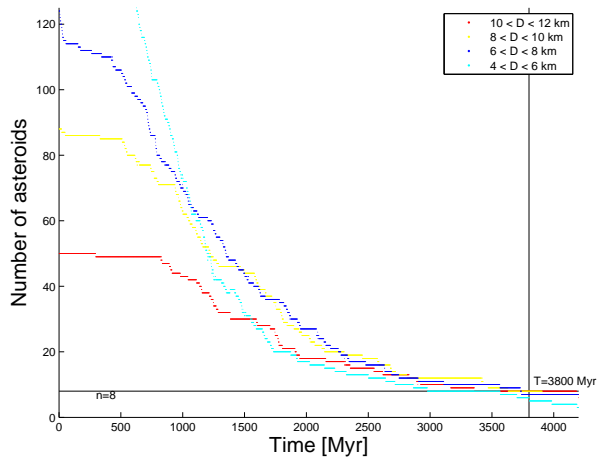}

\caption{Number of Sylvia family members that remained in the Sylvia family
region as a function of time.  See the figure legend for the color code
associated with each given range in particles diameter.}
\label{fig: count_dia}
\end{figure}

In this longer simulation the Sylvia family did not completely dispersed 
up to the end at 4.2 Gyr.  More importantly, 10.0\% of the asteroids with 
$10 < D < 12$~km, 3.4\% of those with $8 < D < 10$~km, 5.6\% of 
those with $6 < D < 8$~km, and 0.8\% of those with $4 < D < 6$~km could
still be found at the end in the Sylvia family region.  These results
our displayed in Fig.~\ref{fig: count_dia}, the horizontal black line
shows the minimum number of objects to identify a cluster in the Cybele
region (8), while the vertical black line shows the minimum estimated
age for the end of the LHB (3.8 Gyr).  See the figure legend for the 
color code associated with the each given range in particles diameter.
As found in the shorter Sylvia simulation, we confirm that the largest
fragments ($D > 7$~km) of a post-LHB Sylvia family could possibly
remain in the Cybele region since its formation.  We identify 
30 X-type objects in this area with $p_v < 0.08$ and $D > 7$~km,
two of which, (149776 and 172973), are current Sylvia family members.
Since space weathering effects on X-type objects are not well understood
(Nesvorn\'{y} et al. 2005), the possibility that some of these asteroids
predates the formation of the current Sylvia family remain speculative
and difficult to prove.  But, it could be an interesting field of 
future research.  Finally, none of the simulated Sylvia fragments
reached the regions of the Ulla and Helga families during 
the 4.2 Gyr simulations, which raises questions about the possible 
mechanisms that put the parent bodies of these groups in their 
current orbits.

Concerning the other integrated families, our results for the Huberta 
family seems to confirm the age estimate obtained in 
Sect.~\ref{sec: chron_huberta}: in both the 1100 and 3000
Myr simulations the family dispersed in timescales comparable with 
the estimated age of the family, i.e., 1100 Myr.  Extrapolating
the current loss rates to the past, we estimated that the Huberta 
family may have been $\simeq 5$ time more numerous at the time 
of its formation.  The upper estimate of the age of 3000 Myr would have
required an initial population more than 10 times larger then the current 
one, which seems unlikely.  Concerning the families proposed by
Vokrouhlick\'{y} et al. (2010), Camilla and Hermione, our results
show that such families would have dispersed in timescales of the order
of 1.5 Gyr at most.  Should such families have formed before that time,
they would no longer be recognizable, as suggested by these authors. 

\section{Sylvia family during the Late Heavy Bombardment}
\label{sec: syl_LHB}

Since results in Sect.~\ref{sec: dyn_cybele} suggested the possibility
that some members of a pre-late heavy bombardment Sylvia family may have
survived to the present, here we investigated what would have happened
to a hypothetical family formed before $\simeq$ 4.0~Gyr when Jupiter 
jumped in the jumping Jupiter scenario of planetary migration of 
Nesvorn\'{y} et al. (2013).  In this work the authors showed that the 
current orbital distribution of Jupiter Trojans may have occurred when 
Jupiter's orbit and its Lagrange points were displaced in a scattering 
event with an ice giant and felt into a region populated by planetesimal.  
Of the three models for this jump capture, we choose to work with their case 1, 
where five giant planets started in the 3:2, 3:2, 2:1, and
3:2 resonant chain, since this scenario was the most favourable in 
reproducing the current orbital distribution of Jupiter's Trojans.

\begin{figure*}

  \centering
  \begin{minipage}[c]{0.5\textwidth}
    \centering \includegraphics[width=3.5in]{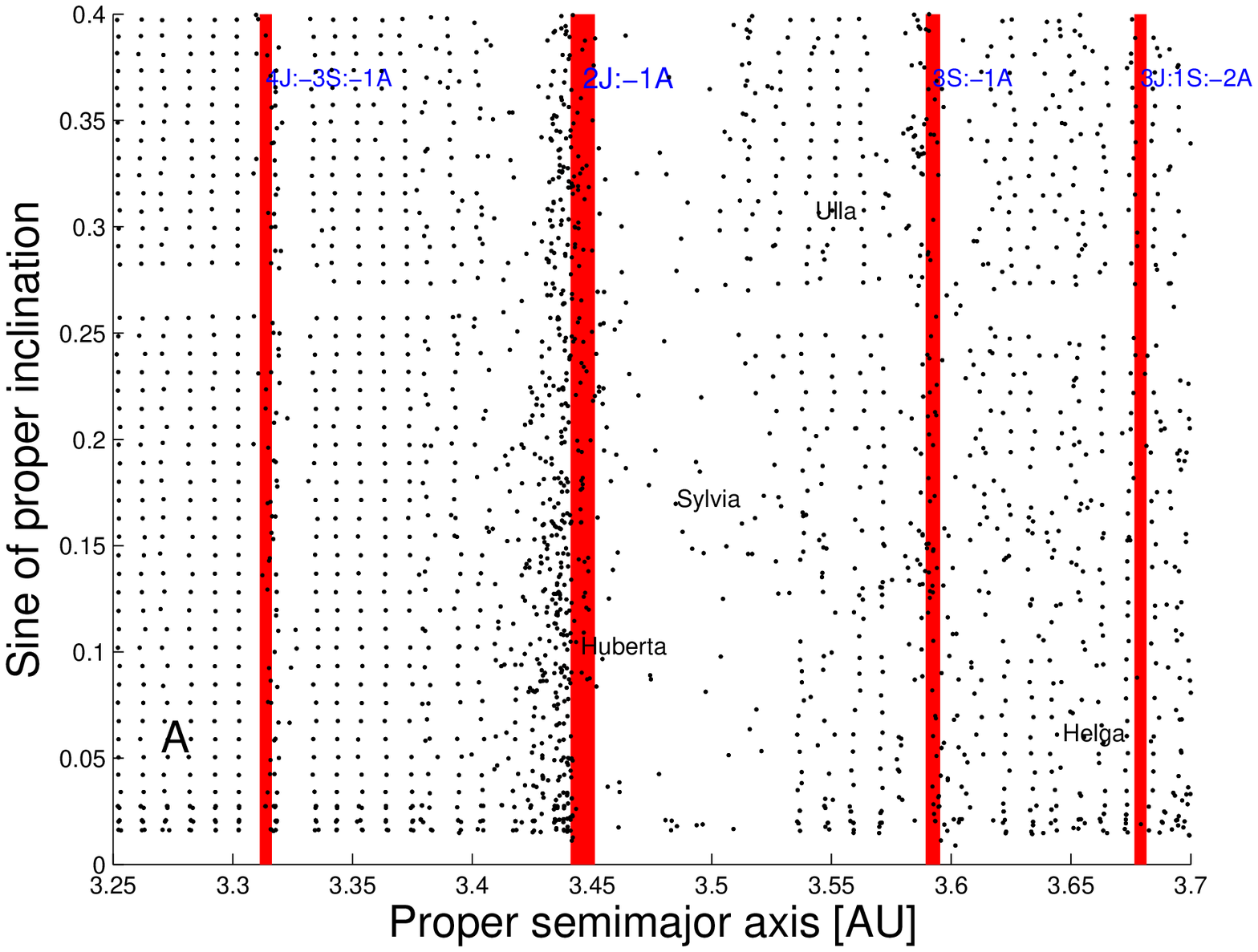}
  \end{minipage}%
  \begin{minipage}[c]{0.5\textwidth}
    \centering \includegraphics[width=3.5in]{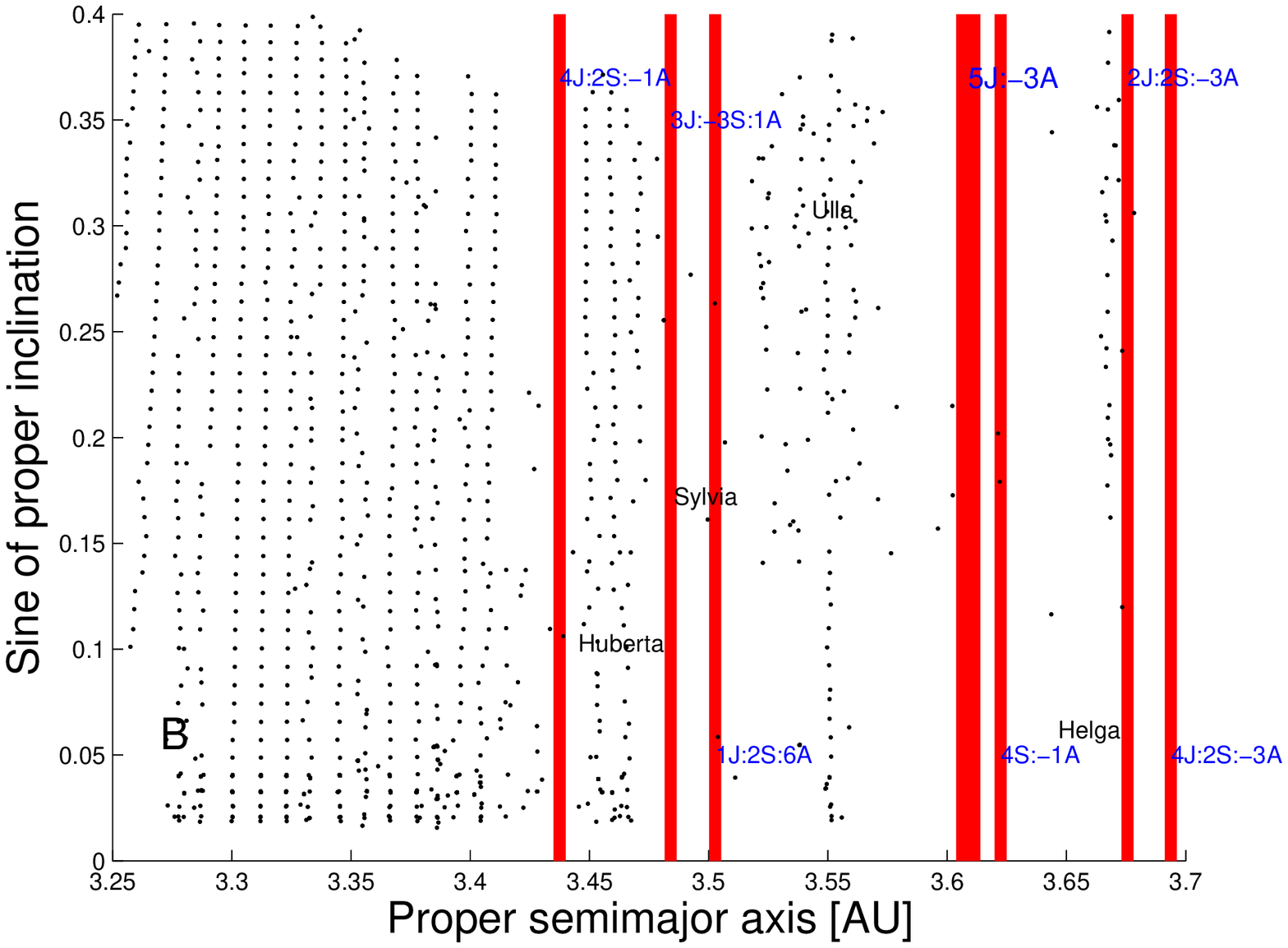}
  \end{minipage}

\caption{A proper $(a,sin(i))$ map of test particles integrated under
the influence of jovian planets before (panel A), and after (panel B) Jupiter
jumped.  Vertical red lines display the position of the main mean-motion
resonances at each epoch.} 
\label{fig: map_jump}
\end{figure*}

To better understand the local dynamics before and after Jupiter jumped, 
we obtained dynamical maps in the $(a,sin(i))$ plane for the planetary 
configurations at the beginning and at the end of the 10 Myr integration
with the scenario of  Nesvorn\'{y} et al. (2013).   We integrated 4131
initially equally displaced particles in the $(a,sin(i))$ plane, in 
a grid of 51 by 81 particles, with a step in $a$ of 0.01~AU and of 
$0.5^{\circ}$ in $i$, starting from 3.20~AU and $0^{\circ}$, respectively.
The eccentricity and the other angles of the test particles were the same 
of 87 Sylvia at present. Fig.~\ref{fig: map_jump} displays our results
for the two simulations.  Black dots represent synthetic proper elements 
computed with the approach described in Sect.~\ref{sec: prop_el}, 
while red vertical lines identify the positions of the main mean-motion
resonances at each epoch.  Mean-motion resonances appear as vertical strips 
deprived of particles, while secular resonances appears as inclined, 
low-density bands.  Since timescales for migration are quite fast in the 
jumping Jupiter scenario, and of the order of the Myr, we ignored 
secular dynamics in this analysis.  One can notice that the current 
Sylvia and Huberta families lie on top of the 2J:-1A mean-motions at the 
beginning of the scenario, and would have been destabilized on timescales 
of less then a few Myr.  After Jupiter jumped, the 2J:-1A mean-motion 
resonance would have migrated towards smaller semi-major axis, but a series 
of three-body resonances currently at higher semi-major axis then the 
center of the 5J:-3A would also have been displaced inwards, making the 
Cybele region, and the Sylvia family region in particular, rather unstable.  
These results suggest that the Cybele asteroids should not be primordial, 
and must have reached their current orbits after Jupiter jumped.

\begin{figure*}

  \centering
  \begin{minipage}[c]{0.5\textwidth}
    \centering \includegraphics[width=3.5in]{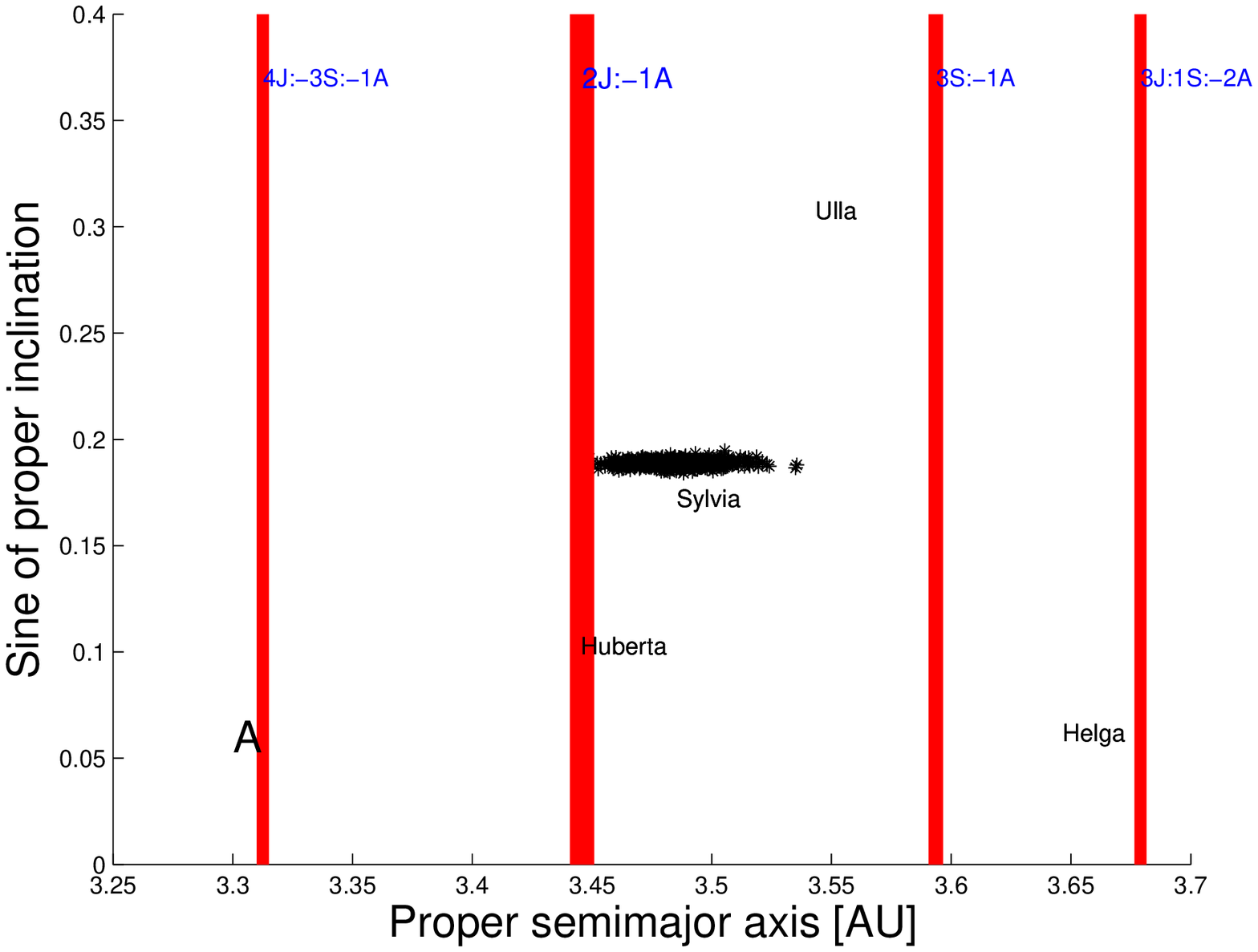}
  \end{minipage}%
  \begin{minipage}[c]{0.5\textwidth}
    \centering \includegraphics[width=3.5in]{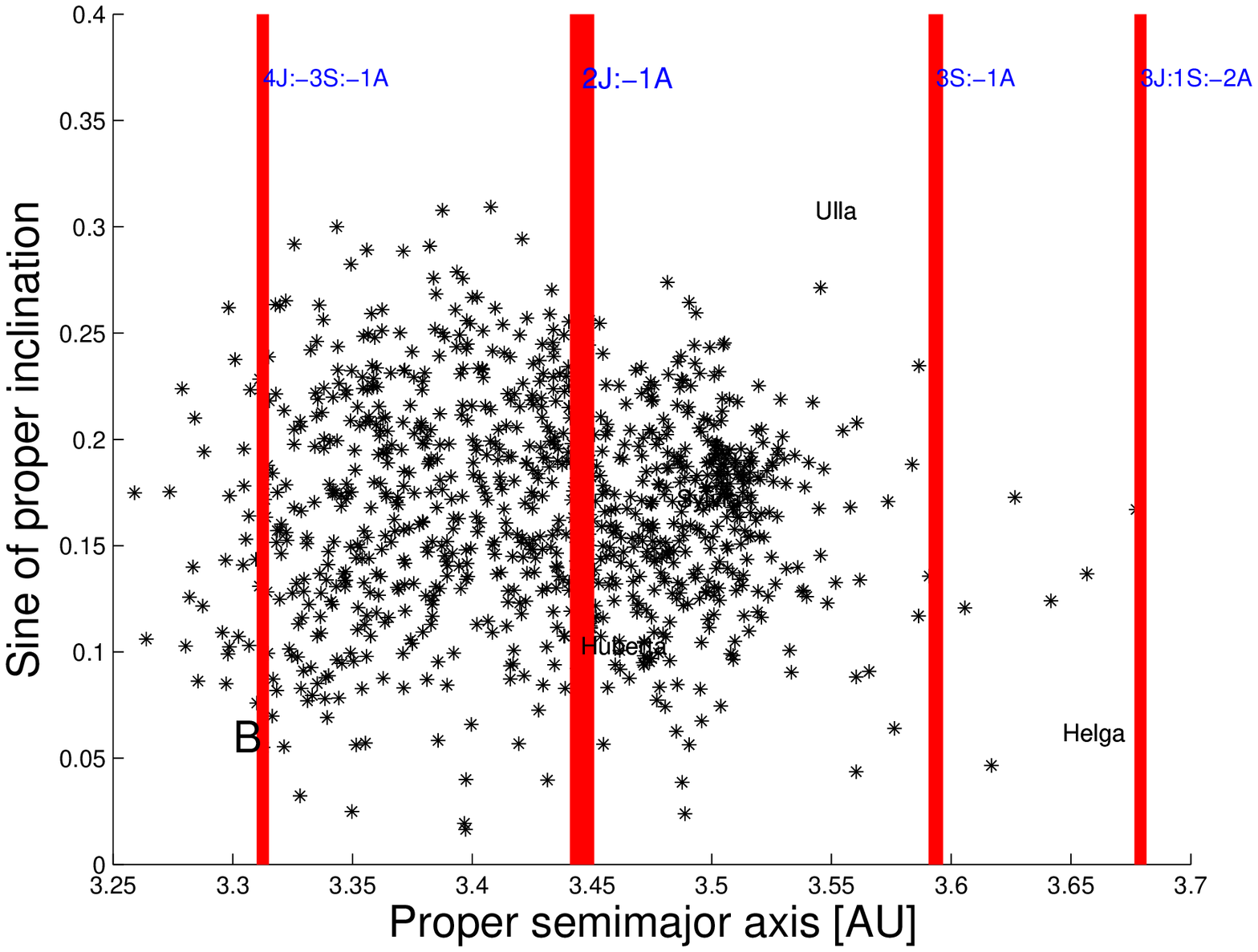}
  \end{minipage}
  \begin{minipage}[c]{0.5\textwidth}
    \centering \includegraphics[width=3.5in]{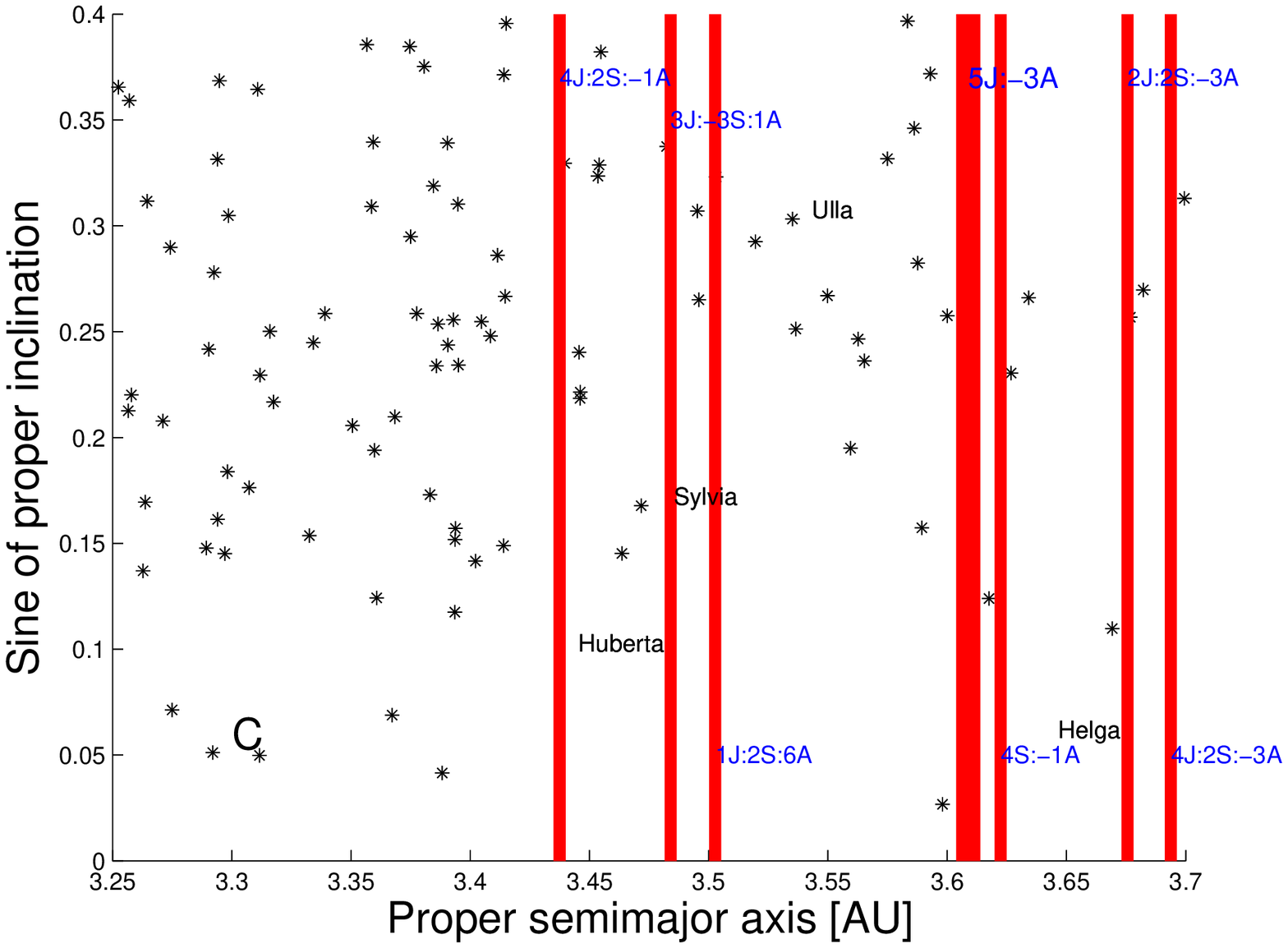}
  \end{minipage}%
  \begin{minipage}[c]{0.5\textwidth}
    \centering \includegraphics[width=3.5in]{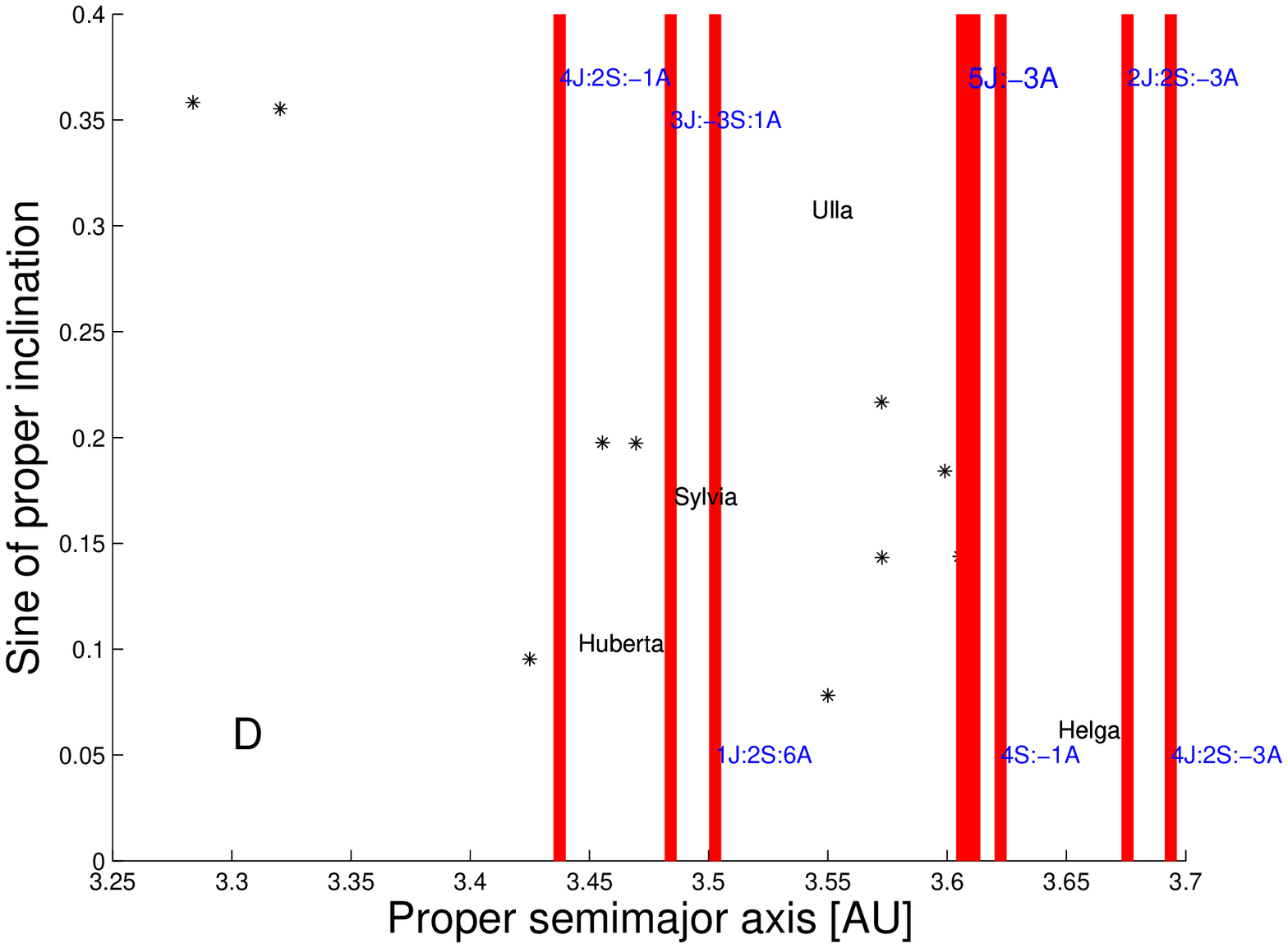}
  \end{minipage}

\caption{Panel A: an osculating $(a,sin(i))$ projection of asteroids at 
the beginning of the simulation with the case I of the jumping Jupiter
model of Nesvorn\'{y} et al. (2013).  Vertical 
lines display the location of mean-motion resonances, 
as in Fig.~\ref{fig: map_jump}, panel A.  Panel B, C, and D: the same 
as panel A, but at t=5, 6, and 10 Myr, respectively.  The instability occurred
at 5.7 Myr. For panels C and D 
we show the location of mean-motion resonances after Jupiter jumped (see 
Fig.~\ref{fig: map_jump}, panel B).} 
\label{fig: sylvia_jj}
\end{figure*}

To confirm the instabilities of asteroid families in the Cybele region
with the jumping Jupiter scenario, we also integrated the same 1000 
particles used for our simulation with the SYSYCE
4200 Myr integration of the Sylvia family discussed in 
Sect.~\ref{sec: past_dyn}, with the approach described in Nesvorn\'{y} et al.
(2013), Sect. 2, over 10 Myr.  We choose to work with this timescales
because this correspond to the period of planetary instability.  In 
this scenario, Jupiter jumped at $t = 5.7$ Myr.  Fig.~\ref{fig: sylvia_jj} 
displays an osculating $(a,sin(i))$ projection of orbital elements of the 
integrated asteroids at the beginning (panel A), and at t =5, 6, and 10 
Myr of the simulation.  Since the particles osculating elements changed 
significantly over the length of the integration, we preferred not to compute
proper elements in this case.  Of the 1000 particles that were simulated 
in this scenario, only 7 had orbits in the Cybele orbital region at the end of
our integration whose orbits were stable enough to allow the computation of
synthetic proper element, i.e., 0.7\% of the initial population.  Overall, 
a pre-LHB Sylvia family would have been almost completely dispersed, so
confirming the results of other authors for the Cybele region
(Bro\v{z} et al. 2013).  Interestingly, 5 particles temporarily reached the Ulla
family region during the length of the simulation, as defined in 
Sect.~\ref{sec: past_dyn}.  Since Minton and Malhotra (2011) argued that
the Ulla region should have been depopulated by the effect of sweeping
secular resonances, and since we showed in Sect.~\ref{sec: past_dyn}
that the parent body of this family may have not reached its current
orbit by mobility caused by the Yarkovsky and YORP effects, our results
suggest that perhaps the Ulla family parent body could have been launched
into its current orbit at the time when Jupiter jumped, either from 
regions at lower semi-major axis, as investigated in this paper,
or, possibly, from the outer main belt.  Investigating the second
scenario is, in our opinion, beyond the purposes of this work, 
that aimed mostly at investigating the dynamics of Cybele asteroids. 
But it could be proposed as an interesting possible future line of 
research.

\section{Conclusions}
\label{sec: conc}

In this work we:

\begin{itemize}

\item Obtained high-quality synthetic proper elements for 1500 among 
numbered and multi-opposition asteroids in the Cybele region and computed
the location of two-body and three-body mean-motion resonances using 
the approach of Gallardo (2014) for all resonances with a strength parameter
$R_S$ up to $10^{-5}$ for two-body and $10^{-4}$ for three-body resonances.
As discussed in Gallardo (2014), the number density of mean-motion resonances
grows considerably for semi-major axis larger than 3.7 AU.

\item Studied the secular dynamics in the Cybele region and identified
the population of librators currently in non-linear secular configurations.
We identified a population of 32 asteroids in $g+s$-type secular resonances,
22 of which in $z_1$ librating configurations, so confirming the important
role that this resonance has in shaping the dynamical evolution of asteroids
in this region (see Vokrouhlick\'{y} et al. 2010).

\item Revised the information on physical properties (taxonomy, photometry, 
and geometric albedos) available in current spectroscopic surveys, 
SDSS-MOC4, and WISE data.  As in Carruba et al. (2013), we found that the 
Cybele region is dominated by dark, low-albedo, primitive objects, with a 
sizeable fraction of D-type bodies that were not detectable with the methods 
used in Carruba et al. (2013).

\item We obtained dynamical families in the area using the Hierarchical 
Clustering Method of Zappal\'{a} et al. (1995), obtained a preliminary
estimate of the family age using the method of the Yarkovsky isolines, 
and eliminated dynamical interlopers, objects with taxonomical properties
consistent with those of other members of the family but whose current
orbits current have been reached by dynamical evolution during the maximum
estimated age of the family, using the approach of Carruba et al. (2014b).
We identified the Sylvia, Huberta, and Ulla families of Nesvorn\'{y} et
al. (2015) and the Helga group of Vinogradova and Shor (2014), and  
confirmed the possibility that the Sylvia family might date from the last
stages of planetary migration. 

\item Used the Monte Carlo approach of Vokrouhlick\'{y} et al. (2006a, b, c),
modified to account for the stochastic YORP effect of Bottke
et al. (2015), and variability of the Solar luminosity, to obtain 
refined estimates of the
Sylvia and Huberta family ages and ejection velocity parameter.
Two possible family ages were obtained for the Sylvia family, 
$T = 1220\pm40~Myr$, and $T = 4220^{+100}_{-120}~Myr$, suggesting the 
possibility that multiple collisional events occurred on this asteroids.
The Huberta family should be $T=1100 \pm 50~Myr$ old.

\item Investigated the past and future dynamical evolution of the Sylvia,
Huberta, and of the possibly currently depleted families of Camilla and 
Hermione using newly developed symplectic integrators that simulates both 
the stochastic YORP effect and variability of the Solar
luminosity.   We confirmed the age estimates for our families found
with previous methods and found that a fraction of about 5\% of the
largest fragments of a hypothetical Sylvia family that formed just 
after the LHB could have remained in the Cybele region up to these days
(but would be of difficult identification).  Any family that formed from 
the binary asteroids Camilla and Hermione should have dispersed in 
time-scale of 1.5 Gyr at most, in agreement with previous results 
(Vokrouhlick\'{y} et al., 2010).

\item Studied the dynamical evolution of a fictitious Sylvia family formed
before Jupiter jumped in the jumping Jupiter scenario of planetary
migration, case I, of Nesvorn\'{y} et al. (2013).   As suggested by other
authors (Bro\v{z} et al. 2013), such family would have been most likely 
dispersed beyond recognition, with only a handful of members surviving
in the Cybele orbital region.  We propose that the parent body of the Ulla 
family may have been scattered to its current orbit during this phase
of planetary migration.

\end{itemize}

Overall, our analysis seems to be in agreement with that of previous works:
we confirm, with some caveats, the possible existence of the Helga group
of Vinogradova and Shor (2014), we confirmed that the Sylvia family should be 
at least 1 Gyr old, with some of its members possibly coming from 
previous, no longer detectable dynamical families.  Families that formed
from the binaries Camilla and Hermione should have dispersed on timescales
of 1.5 Gyr at most (see also Vokrouhlick\'{y} et al., 2010).  Finally, 
any dynamical group that predated the LHB in the Cybele region was most 
likely dispersed when Jupiter jumped.   The possible identification of X-type
asteroids in the Cybele region that formed in possible collisional event 
that predated the formation of the current Sylvia family remains a challenge 
for future works.

\section*{Acknowledgments}
We are grateful to the reviewer of this paper, David Vokrouhlick\'{y},
for comments and suggestions that significantly improved the quality of
this work.  The first author would also like to thank the S\~{a}o Paulo 
State Science Foundation (FAPESP), that supported this work via the grant 
14/06762-2, and the Brazilian National Research Council (CNPq, grant 
312313/2014-4). This publication makes use of data products from the 
Wide-field Infrared Survey Explorer, which is a joint project of the University 
of California, Los Angeles, and the Jet Propulsion Laboratory/California 
Institute of Technology, funded by the National Aeronautics and Space 
Administration.  This publication also makes use of data products 
from NEOWISE, which is a project of the Jet Propulsion 
Laboratory/California Institute of Technology, funded by the Planetary 
Science Division of the National Aeronautics and Space Administration.

\bsp

\label{lastpage}

\end{document}